\documentclass [trackchanges]{aastex62}
\listofchanges

\submitjournal{ApJ}

\begin{document}

\title{Revealing the Two `Horns' of Taurus with \textit{Gaia} DR2}

\correspondingauthor{Graham D. Fleming}

\email{gdfleming@uclan.ac.uk}

https://www.overleaf.com/project/5c86a42f0afdb00df4174248\\
\author[0000-0001-9790-8979]{Graham D. Fleming}
\affiliation{Jeremiah Horrocks Institute, University of Central Lancashire,
Preston, PR1 2HE, UK}

\author{Jason M. Kirk}
\affiliation{Jeremiah Horrocks Institute, University of Central Lancashire,
Preston, PR1 2HE, UK}

\author{Derek Ward-Thompson}
\affiliation{Jeremiah Horrocks Institute, University of Central Lancashire, Preston, PR1 2HE, UK}

\author{Kate M. Pattle}
\affiliation{Institute for Astronomy and Department of Physics, National Tsing Hua University,  No. 101, Section 2, Guangfu Road, Hsinchu 30013, Taiwan}
\affiliation{Centre for Astronomy, School of Physics, National University of Ireland Galway, University Road, Galway, Ireland}

\begin{abstract} 

We investigate the spatial properties of sources from the \textit{Gaia} catalogue previously identified as being members of the Taurus star forming region and which appear in the \textit{Spitzer} catalogue.  We study an area of sky of 10$^\circ\times$15$^\circ$, centred on Right Ascension (2000.0)$=$68.5$^\circ$ and Declination (2000.0)$=$27.0$^\circ$, this being an area surrounding the Taurus molecular cloud. We use data obtained from the \textit{Gaia} DR2 release.  By using an inversion of \textit{Gaia} parallax measurements to obtain distance values and by defining limits to the proper motions of the Taurus moving group, we are able to show that there are substantial differences in depth within the Taurus complex.  Our results suggest that the Taurus cloud has significant depth and that there are two main associations centred at $\sim$130$\pm$6~pc and $\sim$160$\pm$4~pc at {1$\sigma$}. These two associations also have different proper motions, of 24.5$\pm$2.8 and 20.1$\pm$2.4~mas~yr$^{-1}$ respectively. We here label them the `Two Horns' of Taurus.
\end{abstract}


\section{Introduction} 
\label{Intro}

The Taurus molecular cloud (TMC) is one of the closest low-mass star-forming regions, lying at a commonly accepted distance of roughly 140~pc \citep{elias1978study}.  The region covers some 10 to 15~degrees in extent which equates to about 25 to 30~pc at this distance.  This makes comprehensive studies of the entire stellar population of the region difficult and few comprehensive studies of the three-dimensional structure of the cloud complex have previously been conducted (e.g. \citealt{luhman2018stellar}). Situated within the TMC are numerous filaments and smaller cloud structures \citep{hartmann2002flows, schmalzl2010star, kirk2013first, panopoulou201413co, marsh2016census}.  Previous studies have shown that young stars are grouped in and around these smaller structures \citep{gomez1993spatial, kirk2013first}. Early distance measurements \citep{mccuskey1939galactic} determined a distance of 142~pc to the Taurus star-forming region, whilst later studies \citep{straizys1980interstellar, meistas1981interstellar} of a number of Lynds dark clouds in the region \citep{lynds1962catalogue} indicated that the TMC is more extended and exists between about 140 and 175~pc.

A study by \citet{bertout1999revisiting} of three distinct regions of the complex placed the Lynds cloud L1495 at 125.6$^{+21}_{-16}$~pc, the Auriga region at 140$^{+16}_{-13}$~pc, and the southern region at 168$^{+42}_{-28}$~pc.  

The investigation of early-type O and A stars located in the Taurus-Auriga molecular cloud \citep{mooley2013b} within {1$\sigma$} parallax error of 6.2{$<\pi<$}7.8 milli-arcsec (128 to 162~pc), identified a significant number of previously unidentified A5 or earlier stars within the region.  \citet{mooley2013b} also noted in their study that even their new distribution fell far short of the expected number of such stars if a standard log-normal IMF distribution is assumed for the region, adding to the discussion previously noted by \citet{goodwin2004explanation} and other researchers (e.g. \citealt {kraus2017greater}).

In their study, \citet{bertout2006kinematic} derived kinematic parallaxes of 67 members of the Taurus moving group with typical errors of 20\% and identified weak-line and classical T Tauri stars spread over distances between 106$^{+42}_{-24}$ and 259$^{+61}_{-42}$~pc. Very Long Baseline Array (VLBA) parallax observations of the Taurus star-forming regions conducted by \citet{torres2007vlba, torres2009vlba} showed a difference in the distances to separate regions of the Taurus complex by studying a small sample of individual sources. They noted a distance of 161.2$\pm$0.9~pc for the star HP Tau/G2 and 146$\pm$0.6~pc for T~Tau (from \citet{loinard2007vlba}) in the eastern part of the complex, and 130~pc to the central area of the star-forming complex, by observing the T~Tau-type stars Hubble~4 (V* V1023 Tau) at 132$\pm$0.5~pc and HDE 283572 at 128.5$\pm$0.6~pc.  A multi-wavelength photometric study of the Taurus region \citet{gudel2007taurus} using \textit{Spitzer} and \textit{XMM-Newton} data produced large-scale maps detailing the stellar and substellar distribution of the region, and in a similar survey of the TMC conducted by \citet{rebull2010taurus}, pre-main sequence members of the Taurus molecular clouds were identified using the \textit{Spitzer} Space Telescope Taurus project (SSTtau) catalogue (\href{http://cds.u-strasbg.fr/cgi-bin/Dic-Simbad?SSTtau}{http://cds.u-strasbg.fr/cgi-bin/Dic-Simbad?SSTtau}) and Two-Micron All-Sky Survey (2MASS) data (\href{http://vizier.u-strasbg.fr/cgi-bin/VizieR?-source=B/2mass}{http://vizier.u-strasbg.fr/cgi-bin/VizieR?-source=B/2mass}).  

In other, more recent research, planetary-mass brown dwarfs in the Taurus and Perseus star-forming regions have been investigated using photometric and proper motion data from a number of space and ground-based platforms including \textit{Spitzer} and \textit{Gaia} DR1 \citep{esplin2017survey}.

\citet{galli2018gould} presents trigonometric parallax and proper motion observations of Young Stellar Objects (YSOs) in the Taurus region as part of the Gould Belt Distances Survey using the VLBA.  Their data suggest a significant difference between the closest and farthest stars in their sample of about 36~pc with the closest lying at 126.6$\pm$1.7 and the most distant at 162.7$\pm$0.8~pc.  Within this range they noted that the central portion of the L1495 dark cloud is at 129.5$\pm$0.3~pc, whilst the supposedly associated B216 structure lies at 158.1$\pm$1.2~pc. The more recent comparison of \textit{Gaia} DR2 and VLBI astrometry results \citep{galli2019structure} revise these distances but again confirm the existance of significant depth effects within the TMC.

Contemporary studies of the TMC using \textit{Gaia} DR2 data conducted by \citep{luhman2018stellar, esplin2019survey}, present comprehensive studies of the stellar membership of the Taurus region. In both studies extensive reference is made to earlier works with regard to the stellar membership of specific cloud complexes within the TMC and the kinematics of their members. No evidence for an older population of stars previously identified by \citet{kraus2017greater} and \citet{zhang2018pan} is found, however the existence of a possible moving group of stars at a distance of 116 to 127~pc with ages of $\sim$40~Myr first identified in the \textit{Gaia} DR1 data by \citet{oh2017comoving} is suggested.

Previous studies such as those listed above have identified significant distance dispersion among the stellar members of various regions within the TMC star-forming complex, and have further suggested considerable depth effects within the cloud. In contrast to more recent studies, we revisit the coherent catalogue of sources identified by \citet{rebull2010taurus} in the \textit{Spitzer} SSTtau catalogue and use the newly available data from \textit{Gaia} data release 2 to model the characteristics and detailed internal distribution of sources within the region as a whole \citep{brown2018gaia, luhman2018stellar}.  Where our studies overlap, we note that our findings are consistent with those of \citet{luhman2018stellar} and \citet{galli2019structure} and consider them complementary to the results of these studies.

Section~\ref{Gaia} provides a brief overview of the \textit{Gaia} mission and some of the relevant issues concerning DR2. The acquisition of data and its subsequent analysis are discussed in \S\ref{Data} with particular attention being paid to the statistical treatment of the data. Section~\ref{Discussion} presents a discussion of our data which is briefly summarised in \S\ref{Summary}. A compendium of the sources discussed in this study are presented in the Appendix.

\section{\textit{Gaia}} 
\label{Gaia}

The European Space Agency \textit{Gaia} astrometric space observatory \citep{lindegren1996gaia} was launched in December 2013. The spacecraft is designed to measure the parallax, positions and proper motions of stars, with the ambitious goal of producing a three-dimensional map of most of our Galaxy. \textit{Gaia} is not designed to measure distances directly, but they can be inferred through the determination of stellar parallax. The \textit{Gaia} Archive\footnote{\href{http://gea.esac.esa.int/archive/}{http://gea.esac.esa.int/archive/}} is a relational database which can be accessed through an interactive user interface and interrogated using conditional queries.

The second \textit{Gaia} data release (DR2) occurred in April 2018 with a five-parameter astrometric solution for more than 1.33$\times10^9$ sources \citep{brown2018gaia}.  DR2 parallax uncertainties are in the range of up to 0.04~milli-arcseconds~(mas) for sources with a broad-band, white-light magnitude (G) $<$15 and in the order of 0.7~mas at G$=$20. Coupled with proper motion measurements from DR2, a detailed investigation of the internal kinematics of the Taurus star-forming region can be made. Due to the relative proximity of the Taurus star-forming region, where the parallaxes are positive and relative uncertainties are small, a Bayesian prior is not employed in this study \citep{bailer2015estimating, bailer2018estimating, luri2018gaia}, and a straightforward inversion of parallax is used to infer distance. This does not affect any of the conclusions in this paper.

Since it is known that there are unquantifiable (but probably small) parallax errors due to a poorly determined zero-point offset in extinction \citep{lindegren2018gaia}, it is not possible to correct individual \textit{Gaia} parallax values completely. It should also be remembered that \textit{Gaia} is, in essence, an optical telescope and, as such, will have difficulty in accurately measuring parallaxes in areas of high optical extinction due to dust. Hence, for the purposes of this study, mean parameter values are used.  Independent comparisons of \textit{Gaia} and VLBA studies of YSOs in the Ophiuchus, Serpens and Aquila regions \citep{ortiz2018gaia} obtained consistent parallax values across all systems, supporting our use of uncorrected \textit{Gaia} parallax values at this distance.

 \section{DATA} 
 \label{Data}
 
 \subsection{Selection and discussion} 
 \label{DataSelect}
 
Sources towards the Taurus region were selected using a 15$\degr\times$10$\degr$ box centred on RA (2000.0)=68.5$\degr$ and Dec (2000.0)=+27.0$\degr$.  This effectively defines an area on the sky of roughly 126~pc$^2$ at the approximate distance of the Taurus cloud. Parallax values were set between 5.0 and 10.0~mas, setting a box covering a distance range from 100 to 200~pc.  Quality limits were set to only include data with at least 5~independent astrometric measurements. No other quality flags were defined, so as to maximise the number of sources returned, thereby enabling an objective selection of sources within the study area. After performing distance calculations on the mean parallax values of the sources returned, the results were plotted on a histogram in 1~pc bins.  The subsequent distribution showed no evidence of the expected peak in population at around 140~pc. We hypothesised that this was due to the Taurus sources being totally swamped by foreground and background objects. To remove possible contamination of our sample by foreground and background field stars, we cross-referenced our findings with the \textit{Spitzer} SSTtau catalogue \citep{rebull2010taurus} and obtained 168~objects in our search area with known \textit{Gaia} parallaxes and proper motions.

\placefigure{Figures/ParaGmagBDs} 
\begin{figure}[h!]
    \figurenum{1}
    \epsscale{0.5}
    \plotone{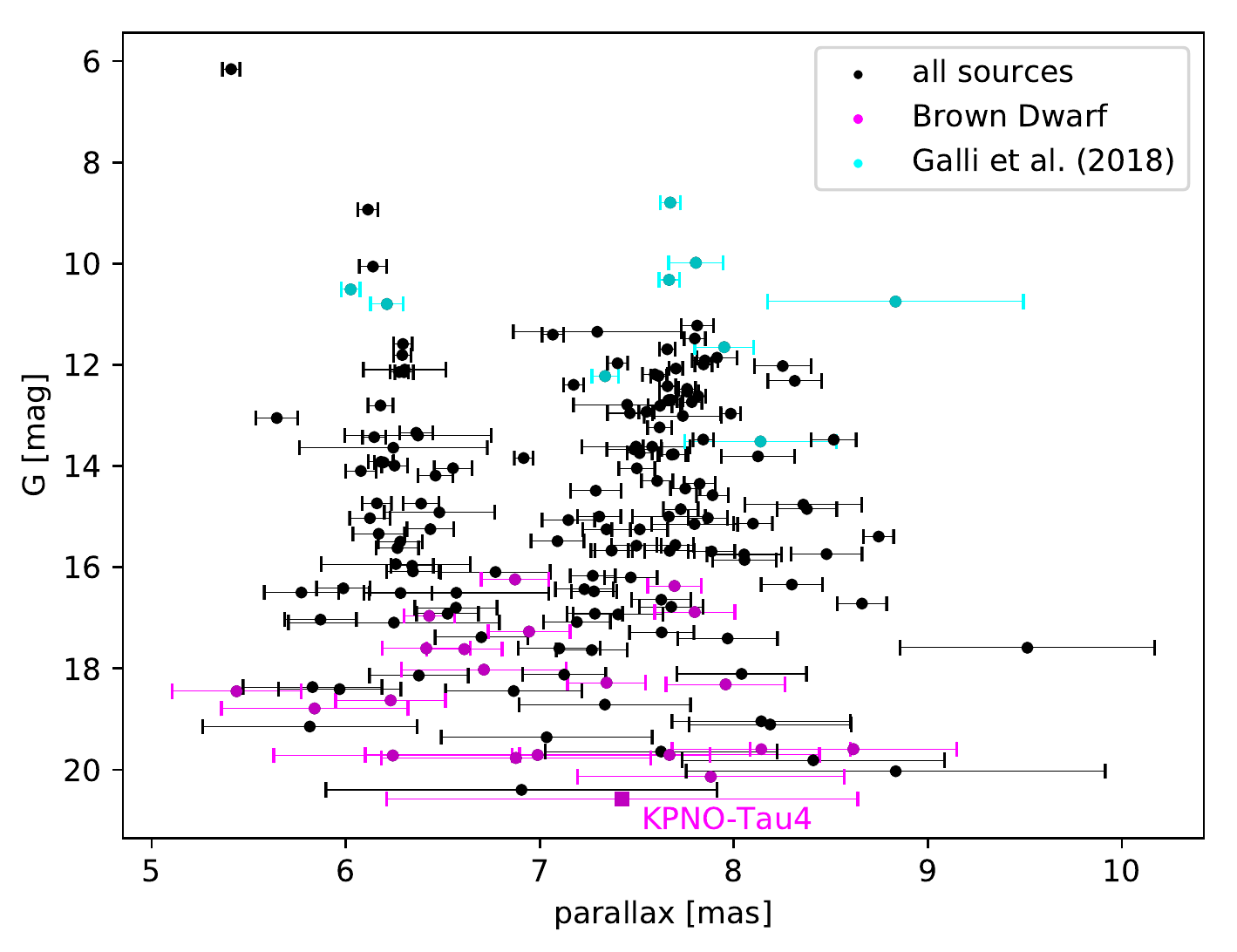}
    \caption{Parallaxes, with errors, of the 168 \textit{Gaia} DR2 sources identified in the \textit{Spitzer} SSTtau catalogue of Taurus members, plotted against G-Band mean magnitude. Sources identified in the SIMBAD Astronomical Database as brown dwarfs are shown in magenta whilst those cross-matched to \citet{galli2018gould} are in cyan.\label{ParaGmagBDs}}
\end{figure}

 Within this subset the largest parallax error is 1.214~mas for object Gaia~DR2~151265002954775936 (KPNO-Tau4), which is a classified L0 brown dwarf (see Figure \ref{ParaGmagBDs}). It should be noted, that brown dwarfs within the sample typically have higher parallax errors, suggesting constraints on the detection of such low luminosity objects. Parallax errors on the remaining sources are significantly lower. Parallax errors in relation to \textit{Gaia} DR2 G-band magnitude (\textit{$phot$\_$g$\_$mean$\_$mag$}) for the sources identified in the \textit{Spitzer} catalogue are presented in Figure \ref{ParaGmagBDs} which clearly shows a double peaked scattering in the parallax (and hence, distance) distribution of the sources. Figure \ref{ParaGmagBDs} also provides a comparison to those of our sources which are present in the \citet{galli2018gould} study and it can be seen that these sources are amongst the brightest of those in our study.

\begin{figure}[h!]
    \figurenum{2}
    \epsscale{0.5}
    \plotone{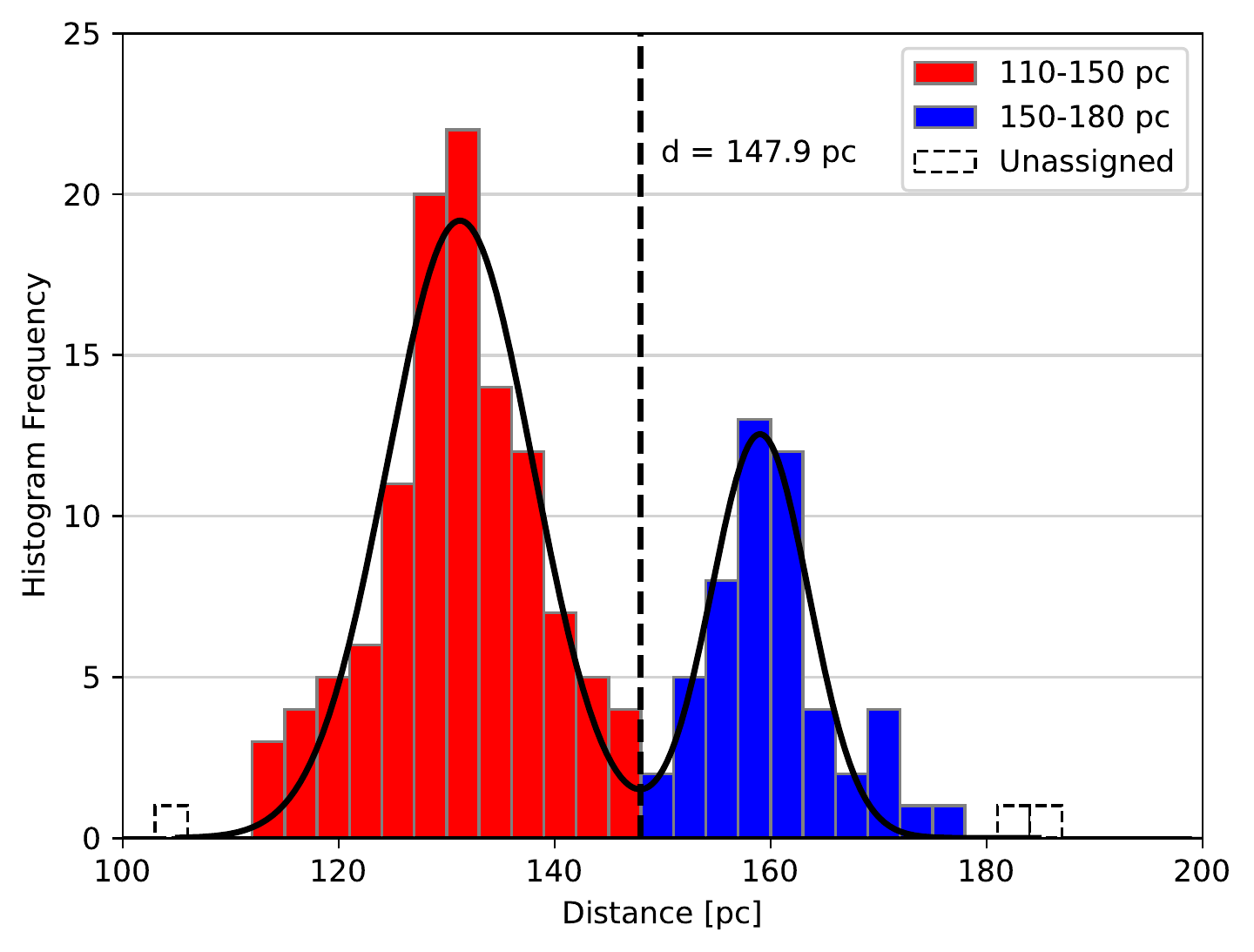}
    \caption{Distance distribution of 168 \textit{Gaia} DR2 sources identified in the \textit{Spitzer} SSTtau catalogue of Taurus members. Sources are grouped in 3-pc bins. Two distinct groupings are seen. A double-Gaussian curve is fitted identifying a minimum value at 147.9~pc (see text for details).\label{Histogram}}
\end{figure}

The properties of the 168~\textit{Gaia} sources are listed in Table~\ref{chartable} in Appendix~\ref{Compendium}. Taking inverse parallax values and determining the distances of our sample of 168 sources, we binned the values at 3~pc intervals and produced a distance distribution.  The resulting distribution is plotted in Figure~\ref{Histogram}. The same double-peaked distribution seen in Figure~\ref{ParaGmagBDs} is visible.

\placefigure{Figures/Histogram} 

A double-Gaussian curve (shown in black) is fitted (Figure~\ref{Histogram}) indicating a minimum in the bi-modal distribution at 147.9~pc, hereafter taken to be $\sim$150~pc.

\subsection{Validation of \textit{Gaia} DR2 distance data} 
\label{ValidDist}

We have compared the \textit{Gaia} DR2 distance data with previous VLBA determinations \citep{torres2009vlba, galli2018gould} to draw comparisons between the two sets of observations and find nine sources common to both studies. These sources are amongst the brightest of our sources and are identified in Figure~\ref{ParaGmagBDs}, showing their \textit{Gaia} G-Band mean magnitudes.

\begin{figure}[h!]
    \figurenum{3}
    \epsscale{0.5}
    \plotone{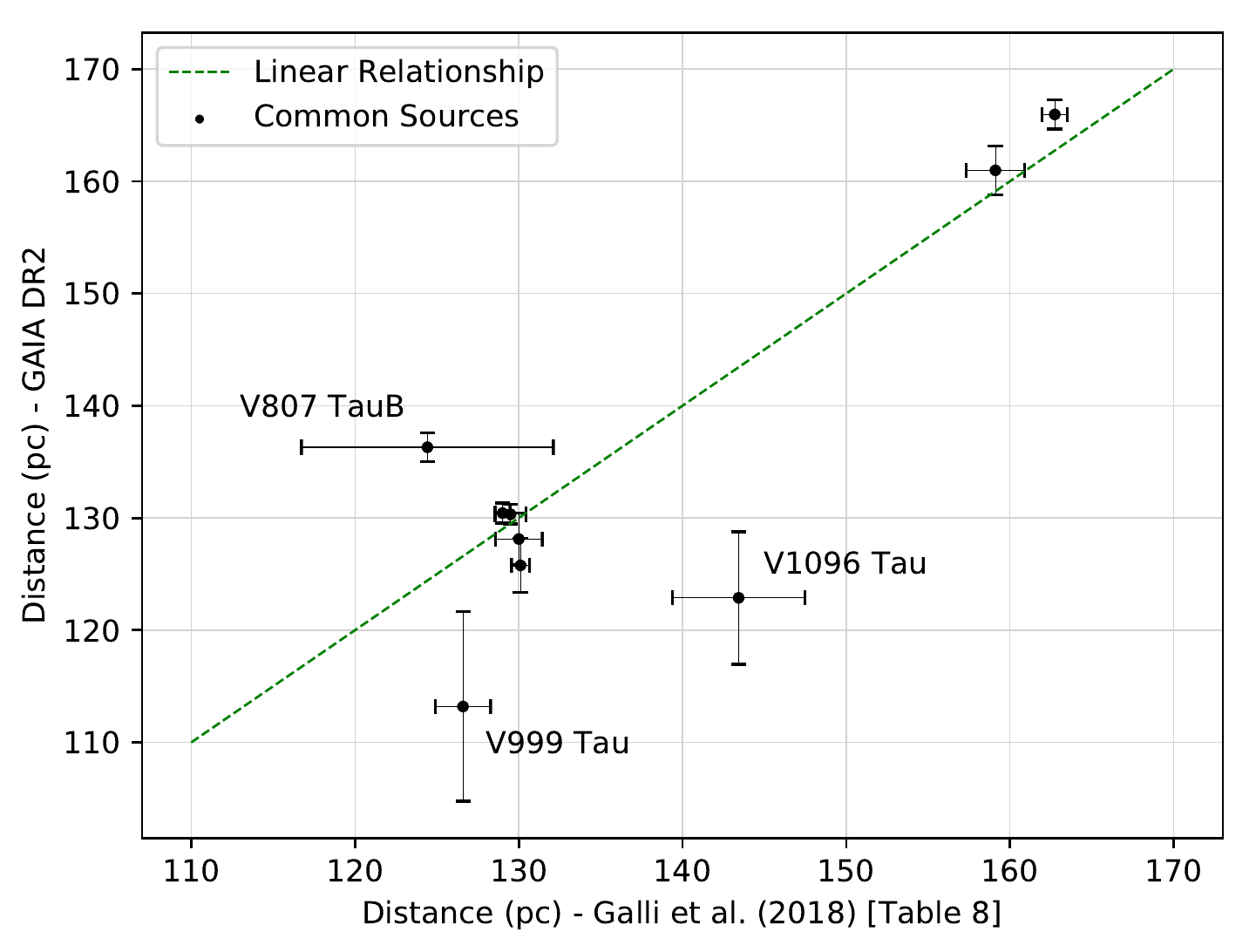}
    \caption{Comparison of \textit{Gaia} DR2 and VLBI \citep{galli2018gould} distance data for the nine common sources.\label{VLBI-Gaia}} 
\end{figure}

Figure~\ref{VLBI-Gaia} plots the two sets of distances of these common sources. It can be seen that that the \textit{Gaia} DR2 derived values are roughly consistent with previous measurements. There are three sources which show large errors, Gaia DR2 147778490237623808 (V807 Tau B), Gaia DR2 148116246425275520 (V999 Tau) and Gaia DR2 163233981593016064 (V1096 Tau). SIMBAD identifies V999 Tau and V1096 Tau as being M0.6 class T-Tauri stars whilst V807 Tau B is listed as a K7 T-Tauri star. \textit{Gaia} DR2 G-band extinction data for V1096 Tau is recorded as being 2.97~mag but is incomplete for the other two sources, it is therefore not possible to make a definitive judgement concerning the large errors displayed by these objects. However, agreement for the other sources is clearly seen.

\placefigure{Figures/VLBI-Gaia.pdf} 

\subsection{Statistical analysis of distance distribution} 
\label{DistStat}

To analyse the distribution of our sources we used Hartigans' dip test for uni-modality \citep{hartigan1985dip, maurus2016skinny}. This is recognised as being a robust statistical measure of the modality of a continuous distribution where the `dip' measures the departure of a distribution from uni-modality. The Hartigan Dip Statistic (HDS), corresponding to the probability `p-value', is determined by repeatedly sampling the maximum difference between the observed distribution of data and that of a uniform distribution that is chosen to minimize this maximum difference. P-values $<$0.05 are an indication of significant bi-modality and values greater than 0.05 but less than 0.10 suggest bi-modality with marginal significance. We obtained a p-value of 0.025. This value suggests that we reject the null hypothesis of uni-modality. We therefore identify our distribution as bi-modal with a boundary between the two groups at $\sim$150~pc.

Having statistically identified that our distance distribution represents two independent populations we split them into `near' and `far' groups using the $\sim$150~pc boundary to give groups lying between 110 to 150~pc and 150 to 180~pc respectively. Figure~\ref{Histogram} identifies our two groups, colour coded red and blue for the `near' and `far' groups respectively. The three sources coded white lie outside the two main populations and are not discussed hereafter. We here identify our two main groups as the `Two Horns' of Taurus. Our analysis is consistent with other recent \textit{Gaia} studies of Taurus \citep{luhman2018stellar, esplin2019survey}.

\subsection{Proper Motion studies} 
\label{PMStudy}

A number of proper motion studies have previously been undertaken of this region, notably those conducted by \citet{jones1979proper, walter1987x, hartmann1991proper, gomez1992ages, frink1997new, ducourant2005pre, bertout2006kinematic} and more recently \citep{dzib2015gould} and \citet{galli2018gould}.  In general, these are all studies of pre-main sequence stars, seeking to identify the proper motions, $\mu$, of members of the Taurus group. Table~\ref{PMstudies} lists the mean, upper and lower proper motion values from these studies. We have used these to constrain the upper and lower limits of proper motion for this study of the TMC.

\begin{deluxetable}{lCCC}[h!] 
\tablecaption{Taurus proper motion values in the literature.\label{PMstudies}}
\tablecolumns{4}
\tablenum{1}
\tablewidth{0pt}
\tablehead{
\colhead{Reference} &
\colhead{$\mu_{min}$} &
\colhead{$\mu_{mean}$} & \colhead{$\mu_{max}$}\\
\colhead{} & \colhead{(mas yr$^{-1}$)} &
\colhead{(mas yr$^{-1}$)} & \colhead{(mas yr$^{-1}$)}
}
\startdata
\citet{jones1979proper} & \nodata & 22.80 & \nodata \\ \citet{frink1997new} & \nodata & 21.24\tablenotemark{a} & \nodata \\ \citet{bertout2006kinematic} & 9.37 & 22.38 & 41.22 \\ \citet{slesnick2006distributed}\tablenotemark{b} & 13.89 & \nodata & 43.05 \\ \citet{torres2009vlba}\tablenotemark{c} &  \nodata & $\sim20.0$ & \nodata \\ \citet{galli2018gould}\tablenotemark{d} & 15.0 & $\sim22.0$ & 39.0 \\
Mean literature values\tablenotemark{e} & 12.75 & 22.14 & 41.09 \\
\tableline
This study & 11.94 & 23.02 & 30.60 \\
\enddata
\tablenotetext{a}{~Value given for the central part of the Taurus-Auriga cloud system.}
\tablenotetext{b}{~Values derived from their figure 9 (lower histogram).}
\tablenotetext{c}{~cited in \citet{dzib2015gould}.}
\tablenotetext{d}{~Maximum and minimum values obtained from their figure 2.}
\tablenotetext{e}{~Ignoring imprecise values from \citet{torres2009vlba} and \citet{galli2018gould}.}
\end{deluxetable}

All 168 identified sources (see Appendix~\ref{Compendium}, Table~\ref{chartable}) have \textit{Gaia} DR2 proper motions, and of these, 165 lie within the `near' and `far' populations mentioned above \textendash{} the remaining 3 are shown in white in Figure~\ref{Histogram}. Based on the literature values given in Table~\ref{PMstudies}, for the purposes of this study, upper and lower limits of proper motion for the Taurus moving group are taken as being 40~mas~yr$^{-1}$ and 12~mas~yr$^{-1}$ respectively. Within these limits there are 161 sources. The four sources which lie outside of our limits have $\mu$ values of 5.41, 9.05, 43.38 and 45.40~mas~yr$^{-1}$ which are consistent with the upper and lower ranges discussed in the literature, but in Appendix~\ref{Compendium}, Table~\ref{chartable} we discounted these $\mu$ values since they are derived using extreme values of either RA or Dec. Discounting these four sources, our minimum, mean and maximum values of proper motion are given in Table~\ref{PMstudies}.

Using a k-Means clustering algorithm we analysed this group to investigate whether there is a proper motion 'split' associated with the distance distribution and to determine the centroids of the groups if such an association exists. We find that such an association does exist and that the mean proper motions of the two groups are different. There are 111 members in the `near' group and 50 in the `far' group \textendash{} the remaining 4 sources are rejected as lying outside of this 12\textendash40~mas~yr$^{-1}$ proper motion cut. The mean proper motion of the `near' population is 24.5$\pm$2.8~mas~yr$^{-1}$, and that of the `far' population is 20.1$\pm$2.4~mas~yr$^{-1}$.

\placefigure{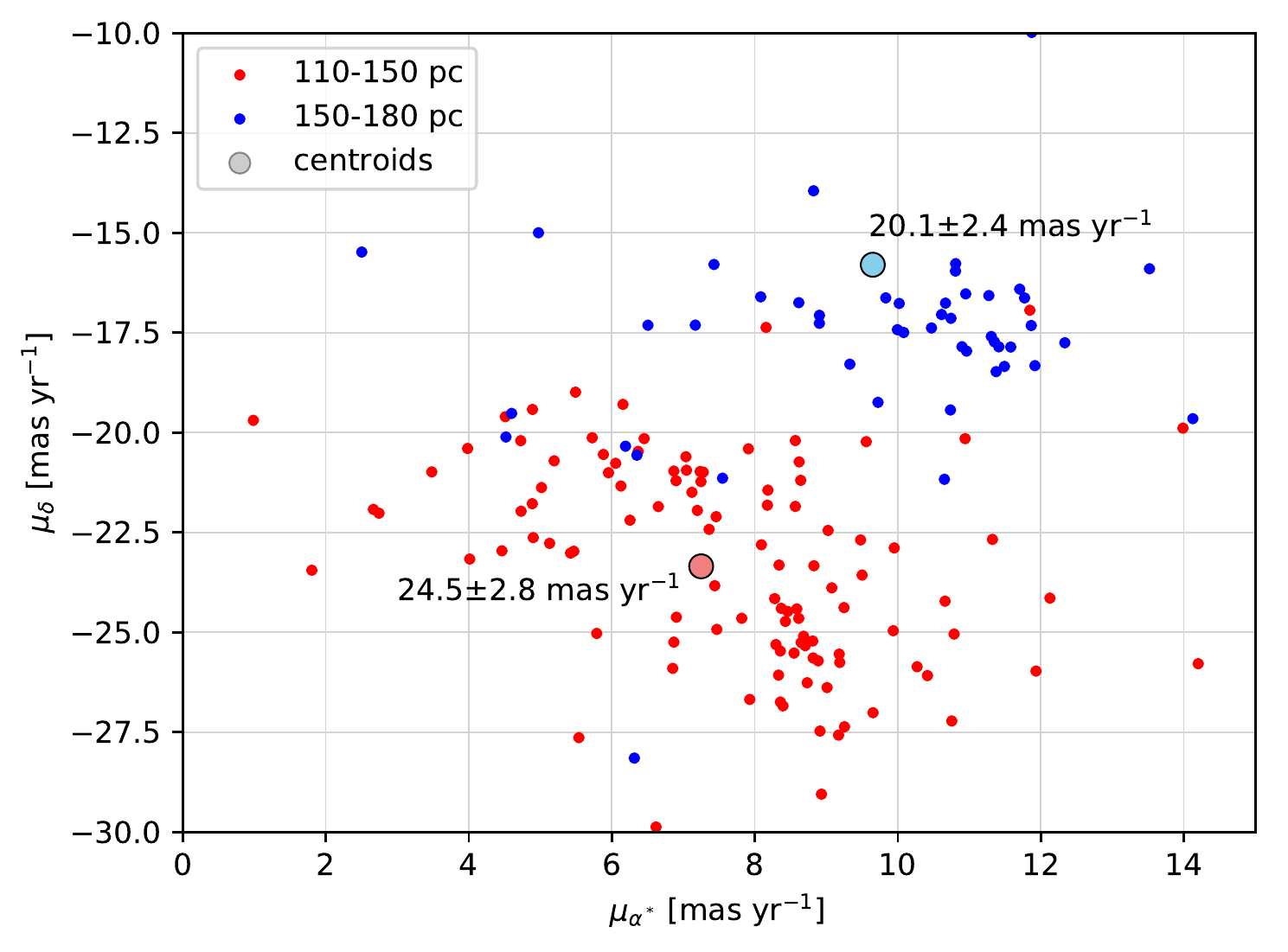} 
\begin{figure}[h!]
    \figurenum{4}
    \epsscale{0.5}
    \plotone{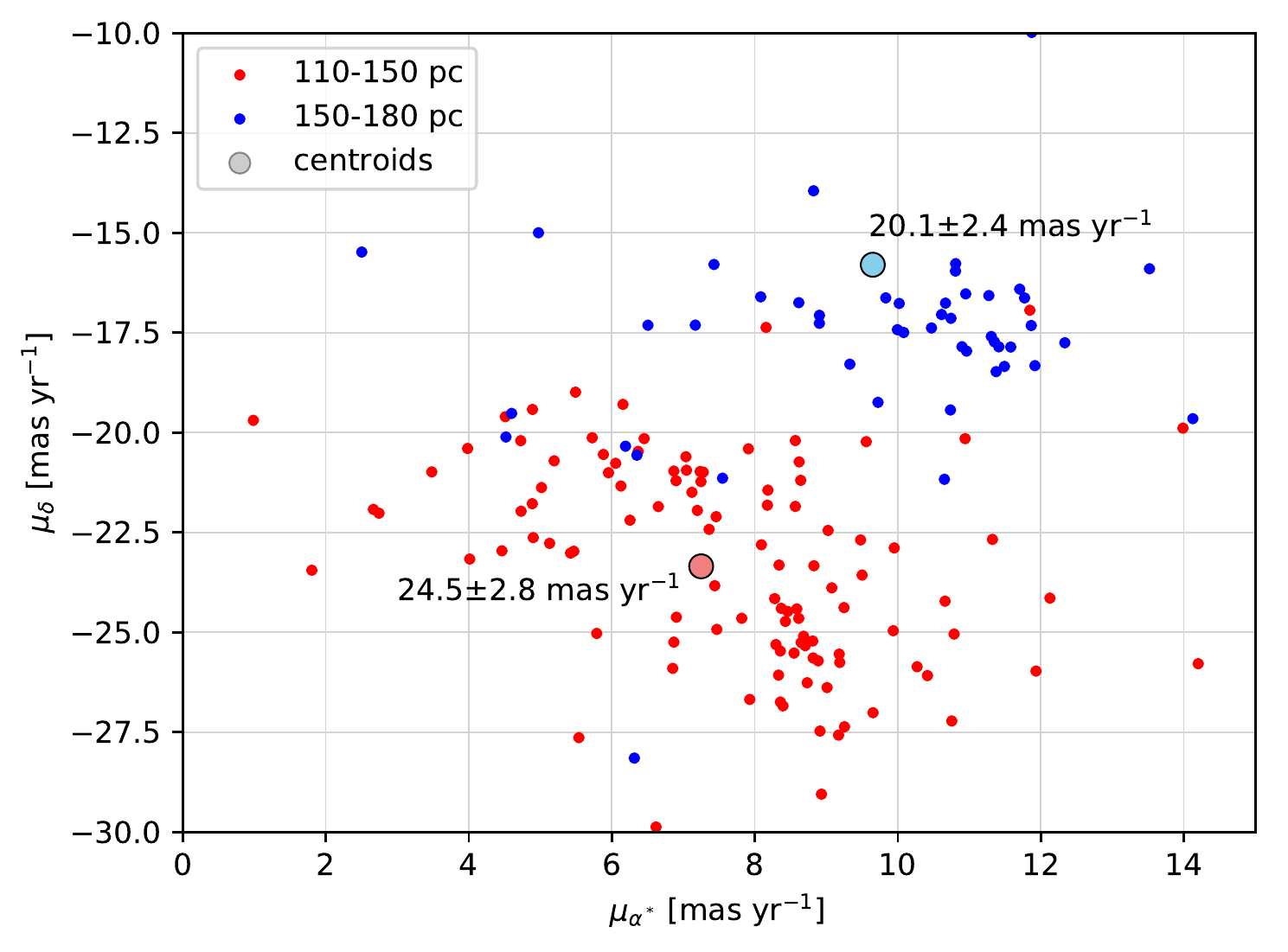}
    \caption{Proper motions for the `near' and `far' populations of 156 of the stars shown in Figure~\ref{Histogram} with proper motion limits of 12~mas~yr$^{-1}$ to 40~mas~yr$^{-1}$. Colour coding is the same as in Figure~\ref{Histogram}. Two populations of proper motion groupings can be seen which are consistent with the two distance groupings seen in Figure~\ref{Histogram}. There are 2 red and 3 blue outlying sources beyond the area shown on this plot (see text for details).\label{PMGroups}}
\end{figure}

The proper motions of 156 of the 161 stars are shown in Figure~\ref{PMGroups} (the colour coding of the two groups is the same as in Figure~\ref{Histogram}). Figure~\ref{PMGroups} has been `zoomed-in' to show the bulk of the sources more clearly \textendash{} so there are five sources lying within our proper motion limits of 12\textendash40~mas~yr$^{-1}$, but which are outside the plotted boundaries of Figure~\ref{PMGroups}. Of these five, two belong to the `near' population (Gaia DR2 146675954953119104 and Gaia DR2 147546080967742720), and three lie in the `far' population (Gaia DR2 148116276529733120, Gaia DR2 147248216395196672 and Gaia DR2 145213192171159552). One of these sources, Gaia DR2 145213192171159552 (CoKu HP Tau G2) has been previously studied \citep{torres2009vlba} using the VBLA, which determined a parallax of 6.2$\pm$0.3~mas. Our \textit{Gaia} DR2 value is 6.02$\pm$0.04~mas and is fully consistent.

For comparison, 7 sources from the study of \citet{galli2018gould} lie within our `near' population and 2 lie in the `far' group. These numbers are statistically low, nevertheless they provide mean proper motions of 24.90$\pm$4.88 and 19.66$\pm$0.50~mas~yr$^{-1}$ for the `near' and `far' populations respectively, which are fully consistent with the values found here. 

\section{Discussion} 
\label{Discussion}

It has been seen (Figure~\ref{Histogram}) that there are two significant peaks in the distance distribution, centred at approximately 130 and 160~pc. Separating these peaks into `near' and `far' populations, as indicated by the red and blue colouring in Figure~\ref{Histogram}, results in mean (and error on the mean) distances for each component of 130.6$\pm$0.7 and 160.2$\pm$0.9~pc respectively. For these groups the standard deviation on the distance is $\sim$6 $\&$ $\sim$4~pc respectively versus a mean error on each measurement of $\sim$4-5~pc. It is probable that the standard deviations for the distances are broadened by these measurement errors. Table~\ref{NearFarStats} shows the parameters of each group. These distributions are roughly consistent with the findings of previous studies mentioned in \S\ref{Intro}, but far more double-peaked than was previously realised.

\begin{deluxetable}{lCCCCCCCcc}[h!] 
\tablecaption{Properties of the near (red) and far (blue) populations shown in 
Figures~\ref{Histogram} and~\ref{PMGroups}.\tablenotemark{a} \label{NearFarStats}}
\tablecolumns{10}
\tablenum{2}
\tablewidth{0pt}
\tablehead{
\colhead{} & \colhead{Number} & \colhead{Mean} &
\colhead{Standard} & 
\colhead{$\mu_\alpha\cos\delta$} & \colhead{$\mu_\delta$} &
\colhead{$\left|\mu_{Total}\right|$} & \colhead{Standard} &
\colhead{Angle} & \colhead{Standard} \\
\colhead{} & \colhead{of Sources} & \colhead{Distance} & \colhead{Deviation} &
\colhead{} & \colhead{} & \colhead{} & \colhead{Deviation} &
\colhead{$\theta$} & \colhead{Deviation} \\
\colhead{} & \colhead{} & \colhead{[pc]} &
\colhead{[1$\sigma$]} & 
\colhead{[mas~yr$^{-1}$]} & \colhead{[mas~yr$^{-1}$]} &
\colhead{[mas~yr$^{-1}$]} & \colhead{[1$\sigma$]} &
\colhead{[degrees]} & \colhead{[1$\sigma$]} 
}
\startdata
Near & 111 & 130.6$\pm$0.7 & 6.7 & 7.5 & -23.1 & 24.5 & 2.8 & 162 & 6 \\
Far & 50 & 160.2$\pm$0.9 & 4.5 & 8.9 & -17.3 & 20.1 & 2.4 & 154 & 17 \\
\enddata
\tablenotetext{a}{~Statistics are calculated after distance and proper motion cuts have been made.}
\end{deluxetable}

For the purposes of simplicity, we have identified our groups as lying at 110\textendash150~pc and 150\textendash180~pc respectively. From Figure~\ref{PMGroups} it can be seen that these two populations have markedly different proper motion characteristics. The populations fall within two separate and distinct proper motion groups, related to their distance.  The mean proper motions of the two groups are listed in Table~\ref{PMstudies}, and are 24.5$\pm$2.8 and 20.1$\pm$2.4~mas~yr$^{-1}$ for the `near' and `far' populations respectively. The mean angles, $\theta$, of the proper motions of the two populations are also listed in Table~\ref{NearFarStats}, along with their standard deviations. These are 162$\pm$6$^{\circ}$ and 154$\pm$17$^{\circ}$ for the `near' and `far' populations respectively, where all angles are measured east of north.

To confirm that our two groups are contained within the same distribution we examined our proper motion groups using a general-purpose non-parametric two-sample Kolmogorov-Smirnov (KS) test \citep{kolmogorov1933sulla}. This two-sample test does not assume that data are taken from Gaussian distributions and is sensitive to differences in both location and shape of the two samples. This test is recognised by \citet{peacock1983two} as being a powerful tool in the analysis of astronomical data. However, we recognise that caution needs to be taken when using this test in astronomical applications \citep[e.g.][]{feigelson2013beware, stephens1974edf}.

The p-value provided by this test can be interpreted in the same way as p-values for other such tests. If the p-value is small, the null hypothesis that the two samples were drawn from the same distribution can be rejected and it can be assumed that the two groups were sampled from populations with different distributions. We obtain a p-value of 4$\times$10$^{-17}$ which supports our earlier identification of a non-unimodal distribution (see \S\ref{DistStat}) and rejects the possibility that the two groups come from the same population.

\subsection{Group and structure correlations} 
\label{GroupStruct}

\begin{figure*}[h!]
    \figurenum{5}
    \epsscale{0.8}
    \plotone{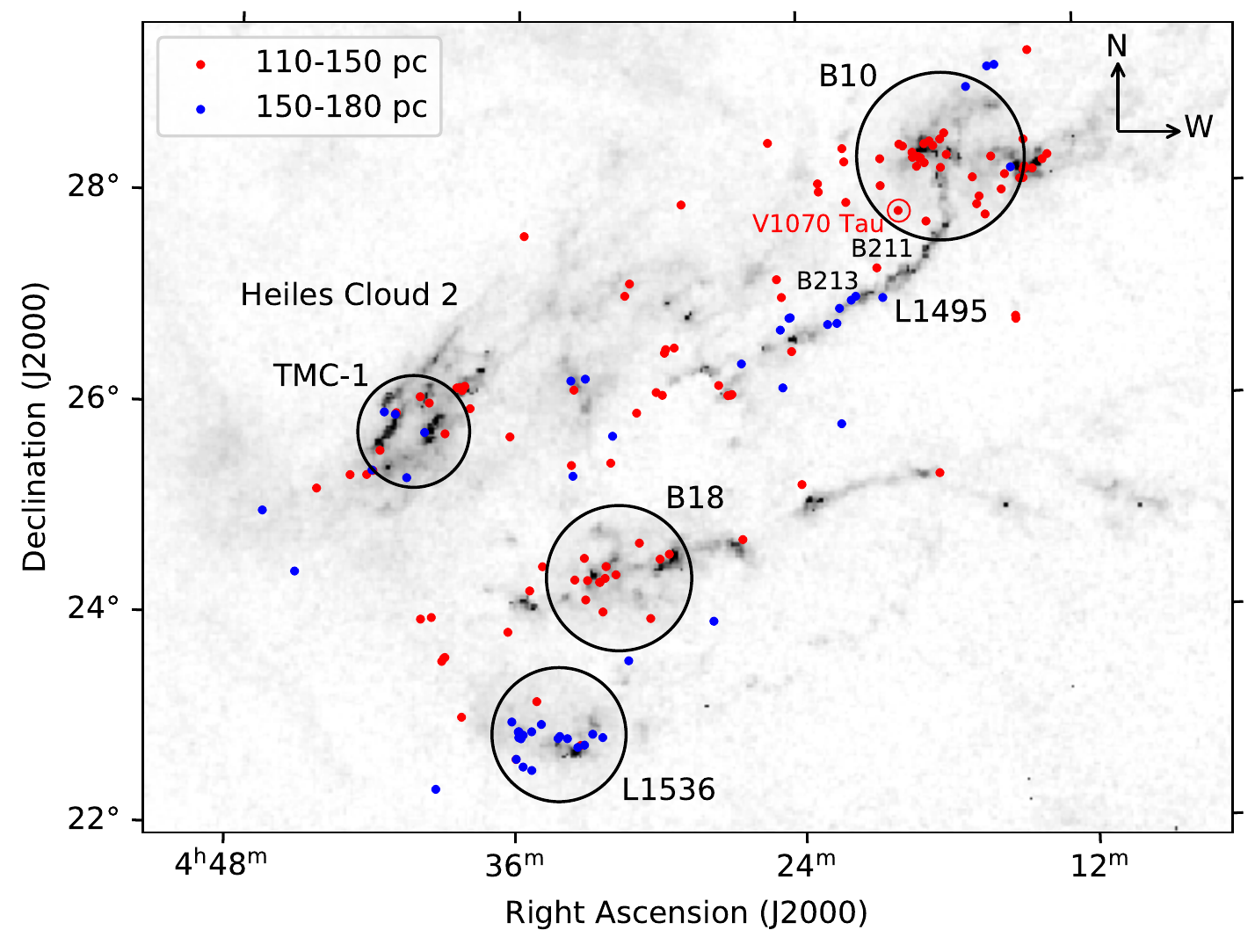}
    \caption{Spatial distribution of \textit{Gaia} DR2 sources identified in the \textit{Spitzer} SSTtau catalogue, using appropriate selection criteria (see text for details). Sources have been identified according to distance and overlaid on a visual extinction map calculated from the 2MASS survey \citep{schneider2011link}.  The colour coding is the same as in Figures~\ref{Histogram} and~\ref{PMGroups}. It can be seen that both B10 and B18 are dominated by sources in the `near' group. L1536 is predominantly composed of sources from the `far' population. See text for further discussion.\label{RedBlueMap}}
\end{figure*}

Using the data presented in Figures~\ref{Histogram} and~\ref{PMGroups} it is now possible to obtain a picture of the distribution of objects within our sample region. Figure~\ref{RedBlueMap} presents this distribution superposed on a visual extinction map of the region calculated from the 2MASS survey \citep{schneider2011link}. Obvious structures within the distance distribution of sources are identifiable within the cloud complex. For example, Gaia DR2 164422961683000320 (V1070 Tau), which lies within the south-eastern region of Barnard's Cloud B10 (part of the extended Lynds L1495 filament), is found to lie at 126.4$\pm$1.6~pc, which is consistent with the value of 126$^{+21}_{-16}$~pc found earlier by \citet{bertout1999revisiting}. The region around B10 can thus be seen to be part of the `near' population, and the 31 sources associated with the `near' group within B10 are found to have a mean distance of 131.9$\pm$3.2~pc, with a standard deviation of 5.0~pc.

\placefigure{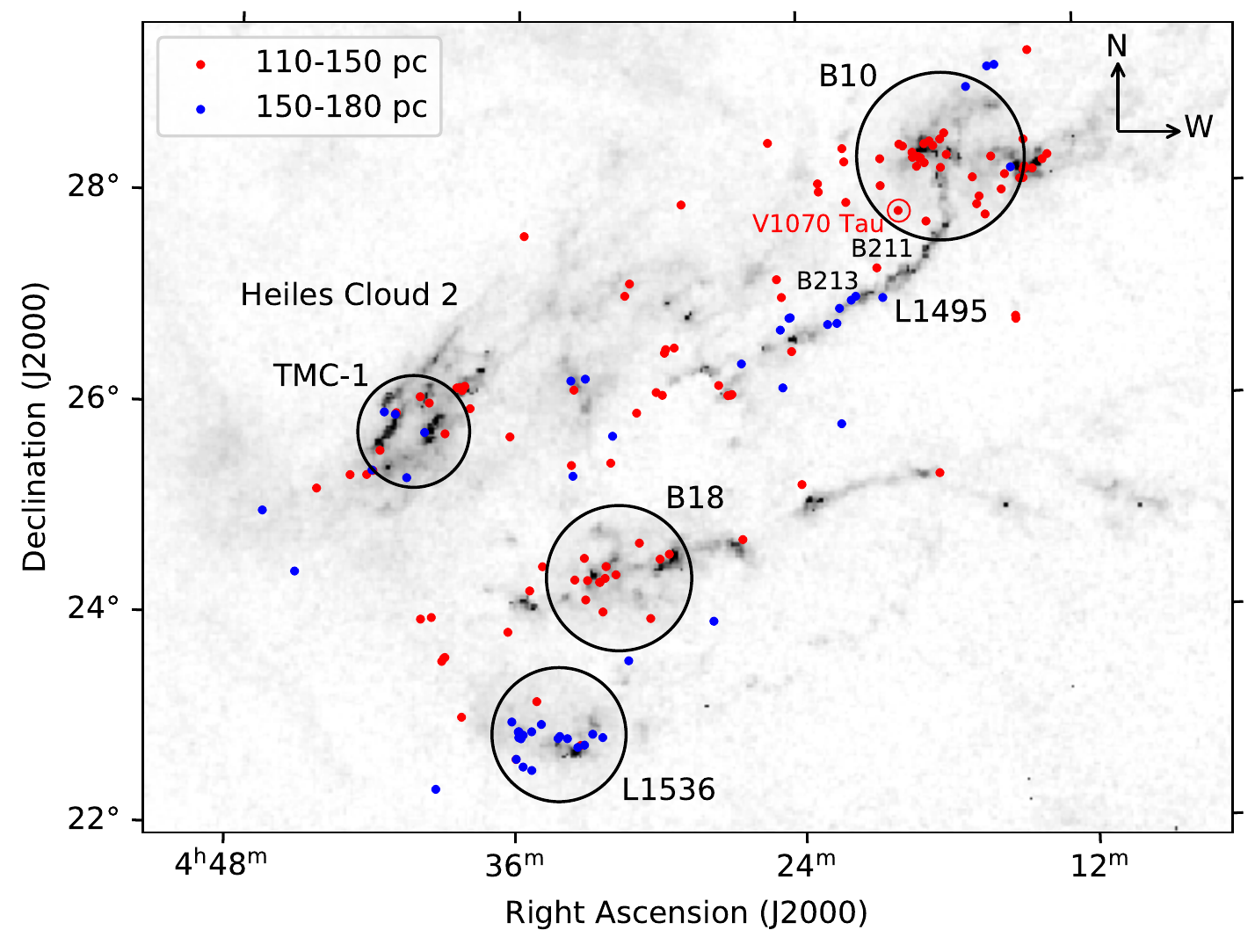} 

Stretching away to the south-east from B10 lies the L1495 filament and its associated clouds B211 and B213. There are a number of sources from both populations which lie directly within, or close to this structure. It is apparent that, if these sources are genuinely associated with the filament, then there appears to be a double distance gradient along this structure. One interpretation of this apparent double gradient is that the cluster of `far' population sources roughly aligned with B213, are actually background to it. For this to be the case, there would need to be gaps in the foreground cloud that allowed the background region to be seen. This explanation would be consistent with the findings of \citet{hacar2013cores}, if one interprets their line-of-sight velocity with distance. This hypothesis is pursued further in \S\ref{DistLOS} below. Cloud B18 appears to be populated with a discrete population belonging to the `near' group.  Analysis of the data for this group shows that they are lying at a mean distance of 127.4$\pm$3.8~pc, with a standard deviation of 7.9~pc.

The VLBA derived parallax to the star HP Tau/G2 in L1536 provides a distance of 161.2$\pm$0.9~pc \citep{torres2009vlba}. HP Tau/G2 also appears in the \citet{galli2018gould} study with a derived mean distance of 162.7$\pm$0.8~pc, which is within 3$\sigma$ of the \textit{Gaia} value of 165.9$\pm$1.3~pc. This star is embedded within the reflection nebula GN~04.32.8, which appears as a crescent-shaped feature in the Herschel column density map of L1536 \citep{kirk2013first}.  HP~Tau/G2 (Gaia DR2 145213192171159552) lies within the area of L1536 (Figure~\ref{RedBlueMap}). There are 20 `far' group sources identified in this area with a mean distance of 160.3$\pm$3.7~pc, and a standard deviation of 6.8~pc. The clear interaction of HP~Tau/G2 with L1536 strongly implies that L1536 is at a comparable distance \citep{kirk2013first}. This is also supported by the earlier study of \citet{bertout1999revisiting}, which placed the southern region of the Taurus cloud at 168$^{+42}_{-28}$~pc. There are seen to be two members of the ‘near’ population situated within, or close to the L1536 region.  These are Gaia DR2 145238687096970496 and Gaia DR2 145157937416226176 which have distance determinations of 130.0$\pm$2.3 and 140.3$\pm$4.2~pc respectively. When considering their maximum distances, they do not fall within the lower boundary of the `far' group and we thus consider their proximity to L1536 as a fortuitous alignment and discount them as being members of L1536.

\begin{figure}[h!]
    \figurenum{6}
    \epsscale{0.8}
    \plotone{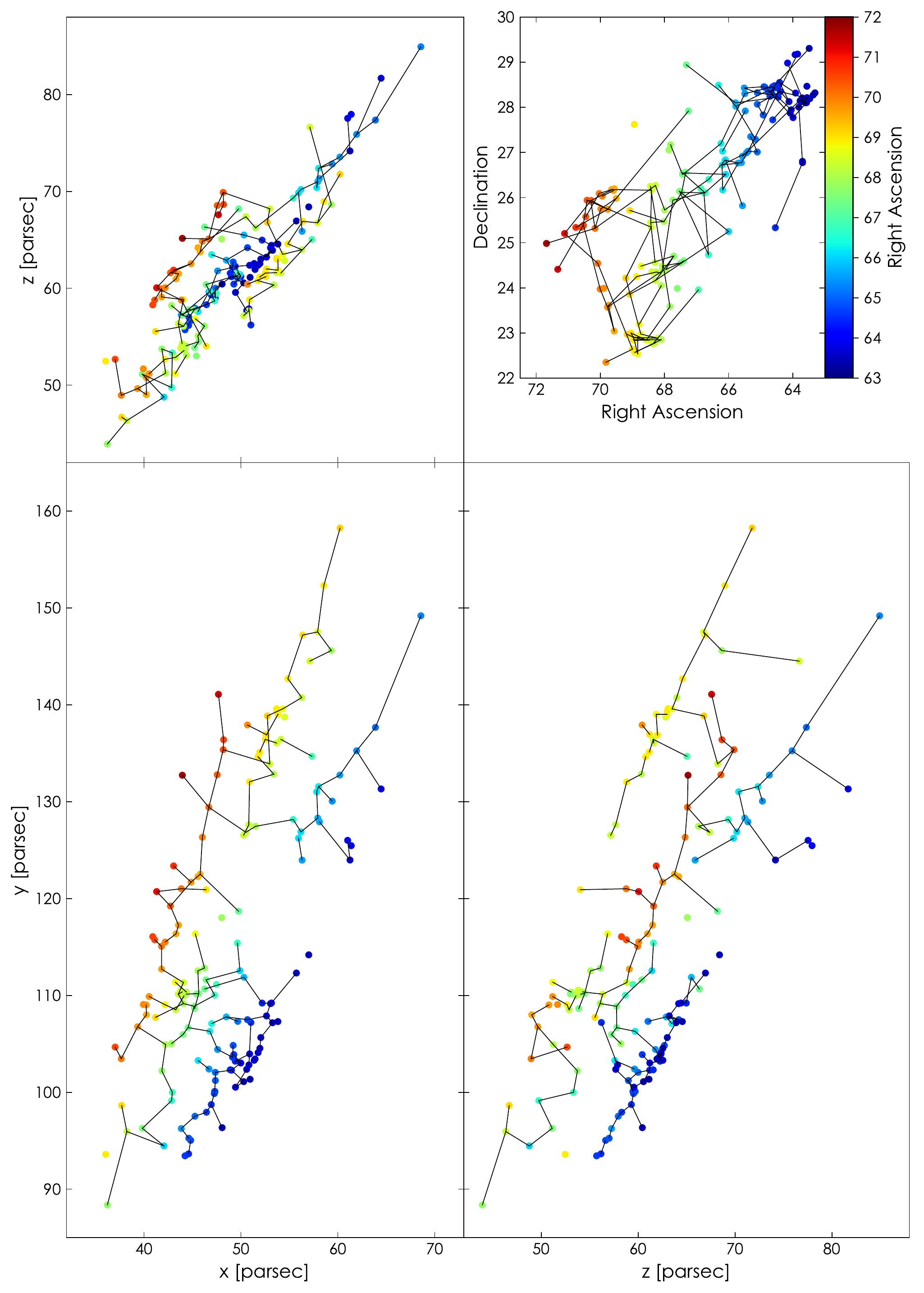}
    \caption{Distribution of 165 sources in the Galactic reference frame. The lower panels suggest that the stellar distribution resembles a 'sheet like' structure (see text).\label{GalXYZ}}
\end{figure}

The association of our groups with the discrete structures within the Taurus cloud is graphically highlighted by Figure~\ref{RedBlueMap}. With the addition of the \textit{Gaia} DR2 derived distance data we are able to develop a three-dimensional picture of this region. So, we studied the 3-D spatial distribution of our sources in Galactic X, Y, and Z as well as celestial ICRS Right Ascension and Declination. Figure~\ref{GalXYZ} presents our 165 sources in the Galactic reference frame ordered by Right Ascension. Of interest here are the lower panels which plot Galactic Z-Y and X-Y with the colour plots ordered by Right Ascension. The `near' and `far' group affiliations previously noted in the ICRS reference frame are clearly seen in Cartesian space with the spatial alignments, suggesting that the TMC is consistent with a structure similar to that of an inclined sheet facing away from us.

\placefigure{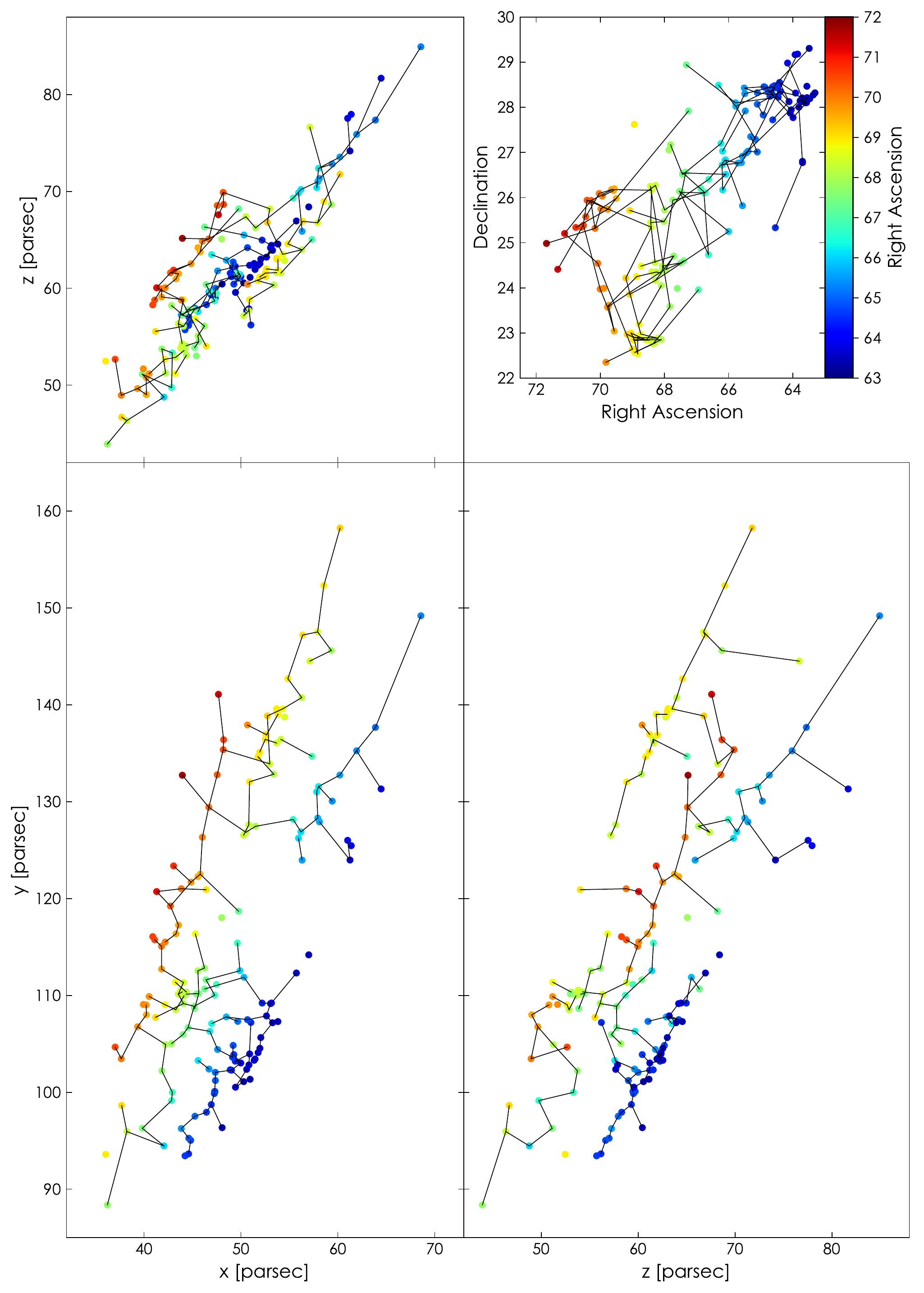} 

\subsection{TMC-1 and the Taurus Molecular `Ring' (TMR)} 
\label{TMR}

Here, we consider the TMC-1 region in more detail. The area commonly known as the `Bull's Tail' \citep{nutter2008scuba} lies within a region previously referred to as the Taurus Molecular `Ring' (TMR), which is associated with the low mass star-forming Heiles~Cloud~2 and has been the subject of many previous investigations \citep[e.g.][]{hartigan2003spectroscopic, toth2004very, nutter2008scuba, malinen2012profiling}. Previous studies of this feature provided a generic distance of $\sim$140~pc in common with the region as a whole. However, \citet{nutter2008scuba} showed that this putative `ring' is not a coherent structure, but rather a chance alignment of discrete sources with different line-of-sight velocities.

Objects Gaia DR2 148401565437820928 and Gaia DR2 148400229703257856 lie in the central region of the `Bull's Tail' and have \textit{Gaia} DR2 derived distances of 136.3$\pm$8.2 and 136.9$\pm$2.1~pc respectively, which are in general agreement with previous studies. From Figure~\ref{RedBlueMap} it can be seen that there is a member of the `far' group to the south of the `Bull's Tail' (Gaia DR2 148374391180009600). The association of members of both the `near' and `far' populations with this feature supports the previous suggestion of \citet{nutter2008scuba} that the TMR is not a coherent feature but is rather composed of disparate sources at different distances spread throughout the depth of the complex.

\placefigure{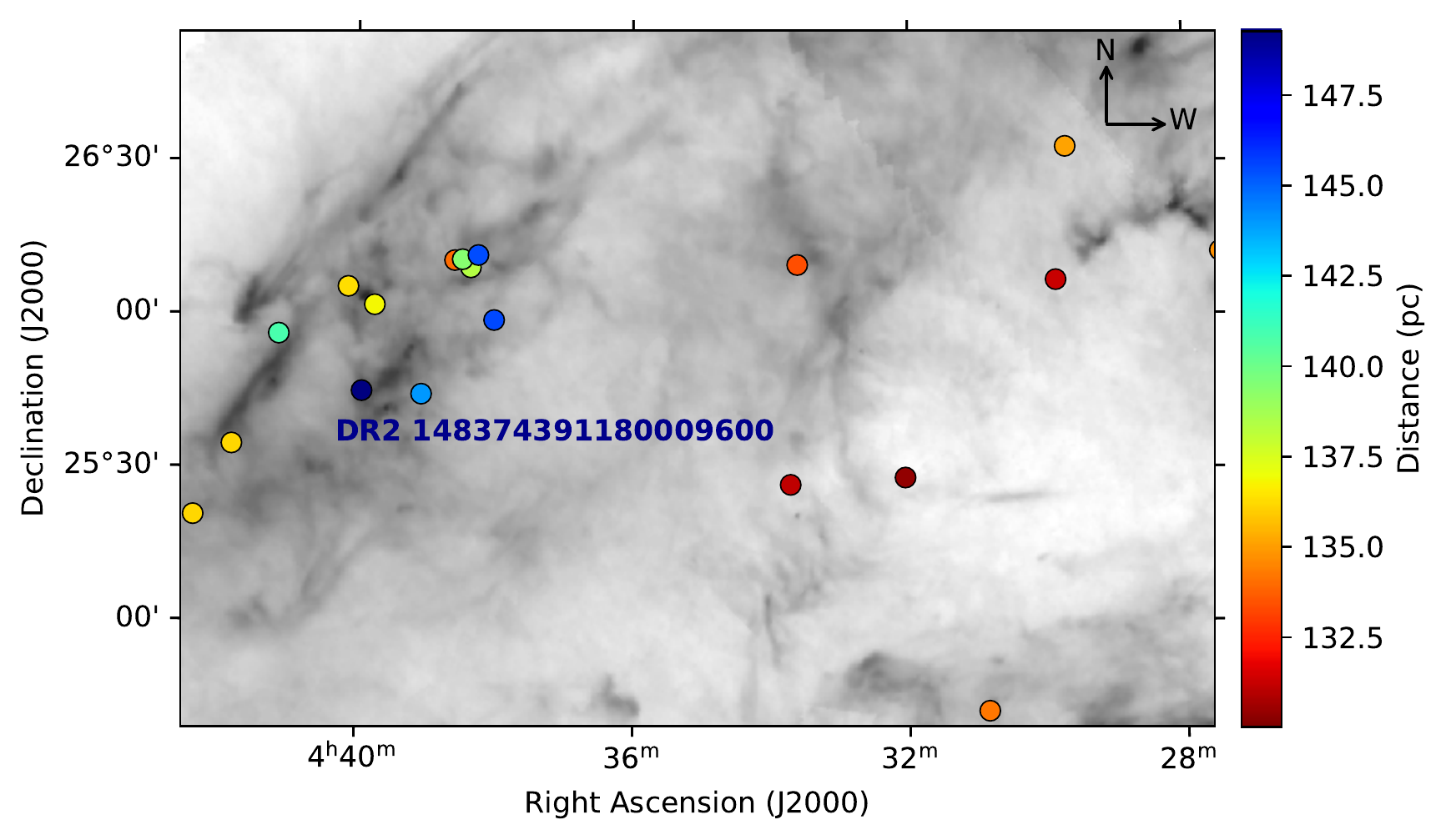} 
\begin{figure}[h!]
    \figurenum{7}
    \epsscale{0.6}
    \plotone{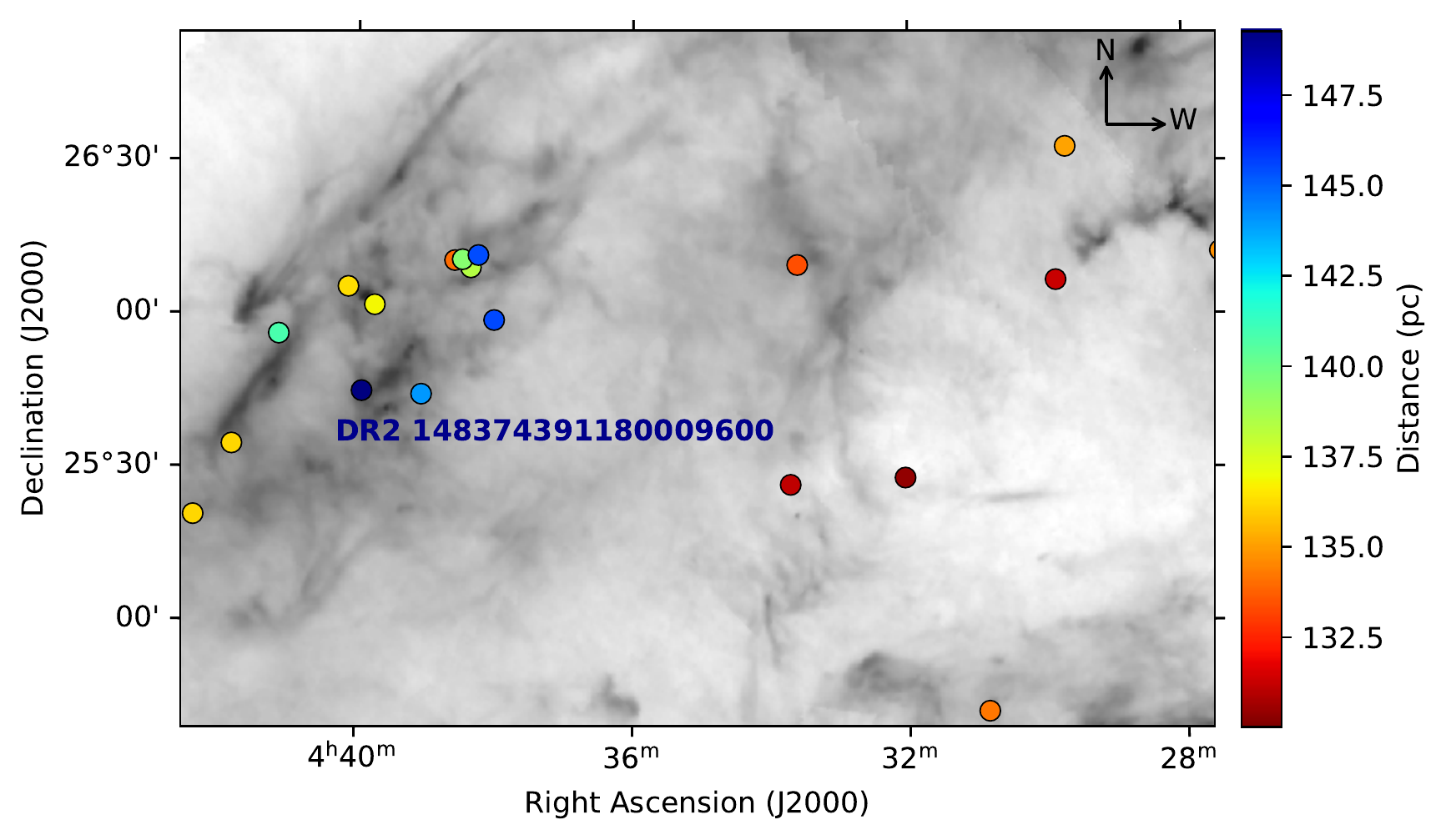}
    \caption{The region around TMC-1 with sources colour coded by distance. The putative `ring' is situated to the east of the figure which clearly shows that the feature has significant depth.\label{TMRDist}}
\end{figure}

Further, the region around the putative `ring' is seen to the east of 4$^h$~36$^m$ in Figure~\ref{TMRDist}. It can clearly be seen that objects in this area have a spread in distance of some 10 to 15~pc. We are thus able to support the hypothesis of \citet{nutter2008scuba} that the so-called `ring' is not a coherent structure.

\subsection{Velocity distributions within the TMC} 
\label{TMCVelocity}

Here we look at the `true' proper motions of our two populations by comparing literature values with our own findings.

\subsubsection{Comparison between distance and line-of-sight velocity} 
\label{DistLOS}

We compare our interpretation of the three-dimensional nature of the TMC with existing line-of-sight velocity measurements of the region. It is well-established that the TMC has a complex velocity structure (e.g. \citealt{clark1977}).  In the context of the \textit{Gaia} distance observations, it is a useful exercise to attempt to associate the stars in our `near' and `far' groups with the major line-of-sight velocity components of the cloud. The $^{12}$CO emission associated with the TMC has systemic line-of-sight velocities ranging from $\sim0-12$\,km\,s$^{-1}$, with the large majority of the emission having velocities in the range $4-8$\,km\,s$^{-1}$ \citep{narayanan2008}.  The TMC has an overall east-to-west velocity gradient, with the eastern parts of the cloud preferentially having a lower systemic velocity than those in the west (e.g. \citealt{goldsmith2008}).  However, there is a great deal of variation within this broad east-to-west trend. Particularly, the L1495 filament is known to have two distinct velocity components, separated by $\sim 1.5$\,km\,s$^{-1}$ (e.g. \citealt{heiles1976}; \citealt{clark1977}).  \citet{hacar2013cores} used C$^{18}$O observations to further separate these two components into multiple sub-filaments, with one set of sub-filaments having velocities $\sim 5-6$\,km\,s$^{-1}$, and the other having velocities $\sim 7$\,km\,s$^{-1}$.  The well-defined plane-of-sky morphology of the L1495 filament is at odds with its apparent lack of velocity coherence, leading to suggestions that the `filament' may in fact be an edge-on sheet (e.g. \citealt{palmeirim2013}).  However, \citet{li2012} compared volume densities derived from dense gas tracers with 2MASS-derived column densities, and found that the high-density portion of the L1495 `filament' has a plane-of-sky depth of only $\sim 0.12$\,pc, suggesting that it is indeed an approximately cylindrical structure.

The stars associated with the TMC included in the \textit{Gaia} DR2 catalogue are located at intermediate visual extinction, and so are not associated with the densest star-forming gas.  We thus compare the distribution of the stars in our two distance groups to the velocities measured in $^{12}$CO observations of the TMC \citep{narayanan2008, goldsmith2008}.  These observations trace moderately dense gas ($n({\rm H}_{2})\sim 10^{2}-10^{3}$\,cm$^{-3}$) which is definitively associated with the TMC, but which is not gravitationally unstable and actively forming stars (e.g. \citealt{difrancesco2007}).

$^{12}$CO velocity channel maps presented by \citet{narayanan2008} show that the B10 and B18 regions have systemic velocities $\sim 7$\,km\,s$^{-1}$, while the L1536 region has a systemic velocity $\sim 5$\,km\,s$^{-1}$.  The L1495 filament shows a double-peaked velocity structure, as discussed above.  The TMC-1 region also has multiple velocities, with some suggestion that the eastern side of TMC-1 is at a lower systemic velocity ($\sim 5-6$\,km\,s$^{-1}$) than the western side (at $\sim 7$\,km\,s$^{-1}$).

We find a correspondence between these behaviours and the spatial distribution of the stars in our `near' and `far' groups (see Figure~\ref{RedBlueMap}).  B10 and B18 are both dominated by `near' stars, and have a systemic velocity of $\sim 7$\,km\,s$^{-1}$, while L1536, containing `far' stars, has a systemic velocity $\sim 5$\,km\,s$^{-1}$.  The L1495 filament, with its two velocity components, contains stars from both groups, as does TMC-1.  However, in TMC-1 the `far' stars are preferentially located in the east, while the `near' stars are preferentially located in the west, corresponding to a velocity gradient from $\sim 5 - 7$\,km\,s$^{-1}$ across the region.  There is thus a qualitative tendency for `near' stars to be associated with $\sim 5$\,km\,s$^{-1}$ sight-lines, and for `far' stars to be associated with $\sim 7$\,km\,s$^{-1}$ sight-lines.

Our results thus tentatively suggest that the two main velocity components of the gas in the TMC are located at different line-of-sight distances, with the $\sim 5$\,km\,s$^{-1}$ gas being located in front of the $\sim 7$\,km\,s$^{-1}$ gas. The `gaps' in the L1495 filament hypothesised in Section~\ref{GroupStruct} are also seen in the velocity data.

\subsubsection{Velocity directions} 
\label{TMCVelVect}

\begin{figure}[!h]
    \figurenum{8}
    \plottwo{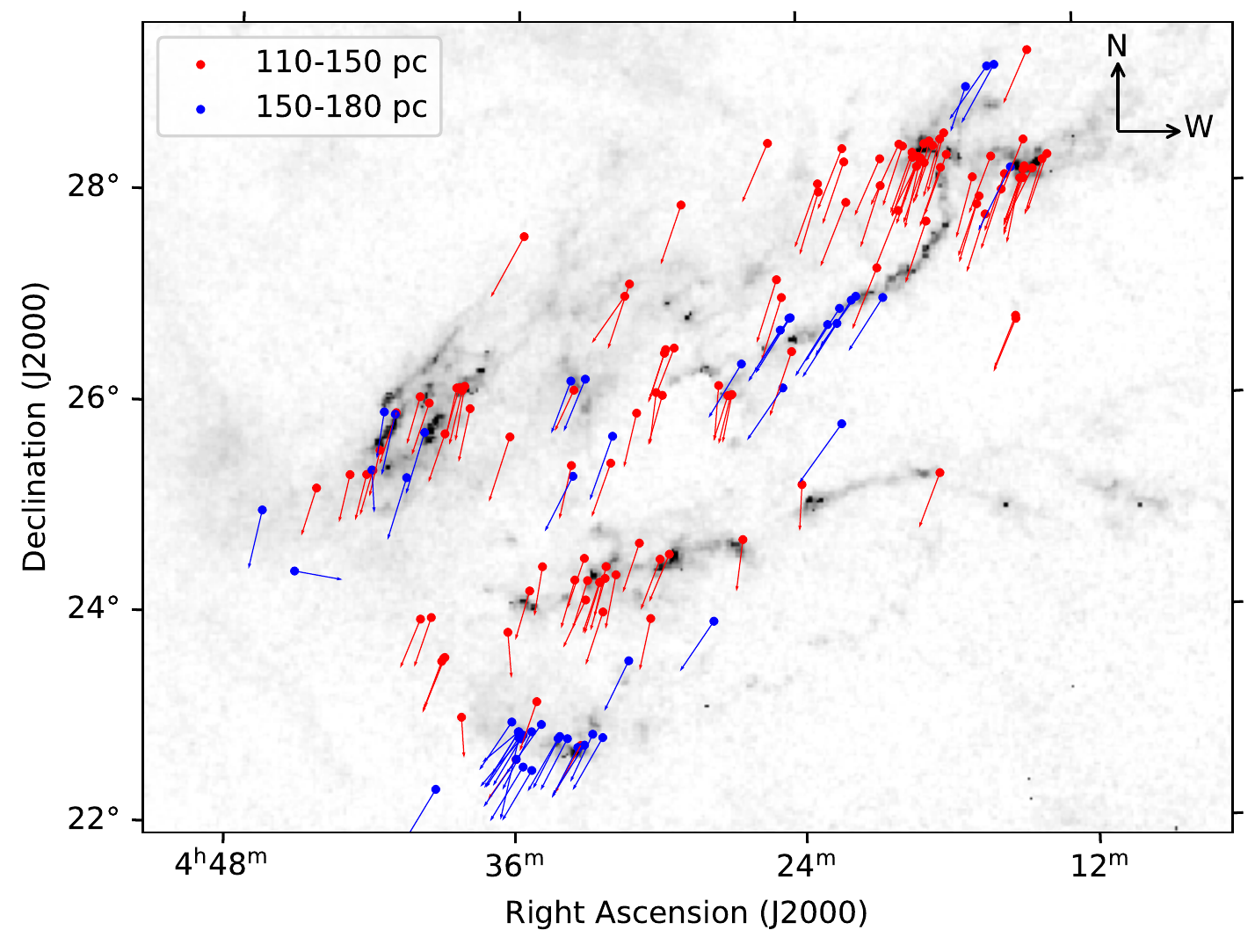}{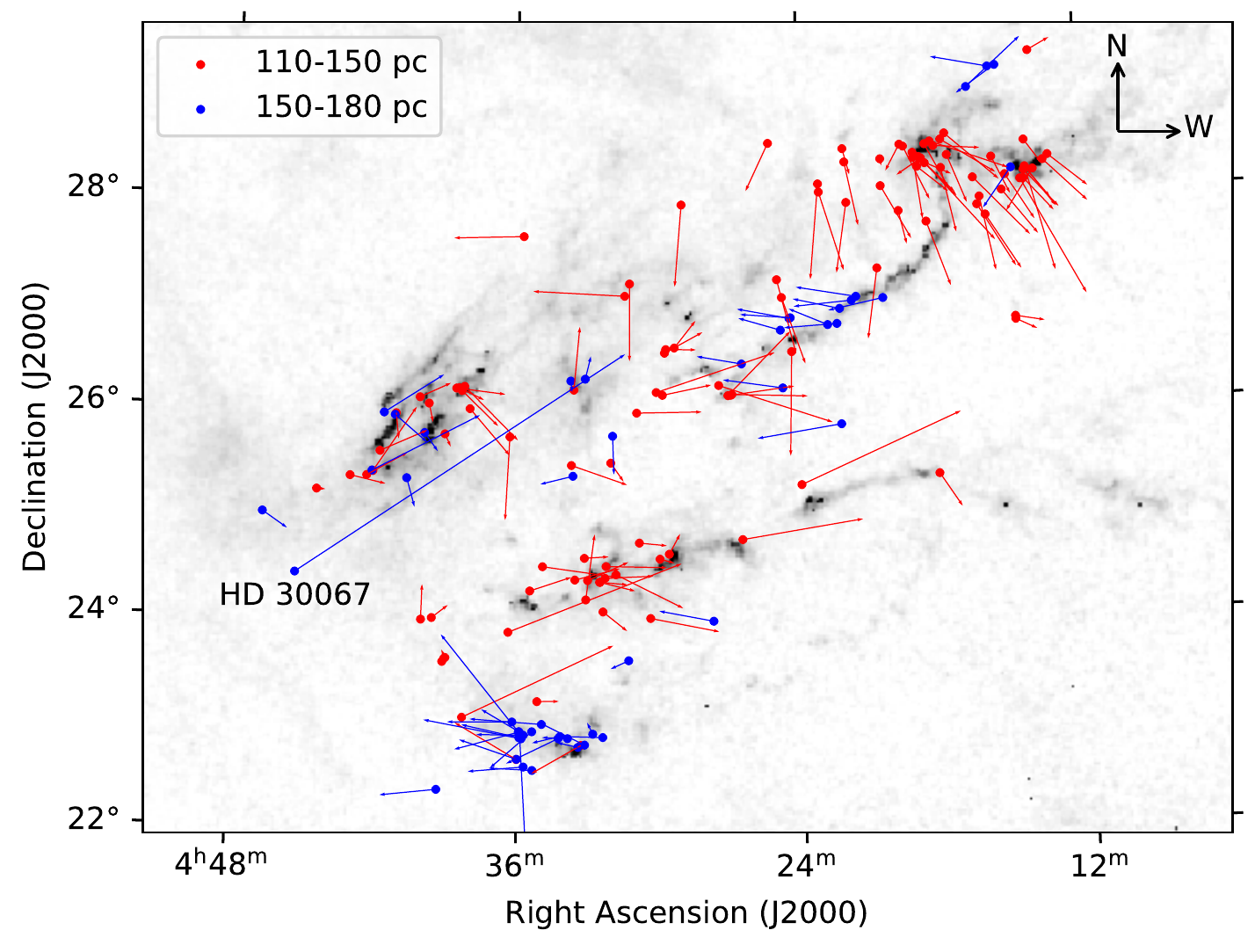}
    \caption{Group distributions within the TMC and related velocity vectors. Left: Proper motions of the `near' and `far' groups identified in Figure~\ref{Histogram} (see text). Individual star proper motions displayed as vectors, showing direction and relative magnitude of velocity. Right: Proper motions after the removal of Solar and Galactic motion components towards Taurus.\label{VelVect}}
\end{figure}

The Right Ascension and Declination proper motions of our sample sources are detailed in Table~\ref{Compendium}. When these values are translated into vectors, as shown in Figure~\ref{VelVect}~(left), it can be seen that there is a small difference in the vectors between the two populations. The mean values of proper motion for each population and their standard deviations are given in Table~\ref{NearFarStats}. The arrows indicate the proper motions of the stars.

However, for a proper analysis of the relative proper motions the influence on these motions caused by the solar motion and Galactic rotation need to be taken into account. Literature studies suggest that this transform is particularly sensitive to the Oort constants \citep{oort1927observational, olling2003oort}, in particular the 'V' component of the solar motion relative to the Local Standard of Rest, which is unknown by up to a factor~2.

\placefigure{Figures/VectorVelocities} 

We use the value of Oort constants from \citet{li2012} and the Solar velocities from \citet{schonrich2010local} using the convention for solar velocities in the Galactic coordinate system as: U being the component toward the Galactic Center; V the component along the line of Galactic rotation; and W being the component out of the Plane, towards the Galactic North Pole.

Since it is also necessary to use a rotation matrix to transform between celestial ICRS (RA,Dec) and Galactic (l,b) coordinates we use the technique presented by \citet{li2019galactic}. We also consider the treatment of barycentric stellar motion in astrometric and radial velocity data. The rigorous treatment of the epoch propagation, including the effects of light-travel time, was developed by \citet{butkevich2014rigorous}. However, for the propagation of the prior information to the \textit{Gaia} reference epoch, it is sufficient to use the simplified treatment, which was employed in the reduction procedures used to construct the Hipparcos and Tycho catalogues, since the light-time effects are negligible at milliarcsecond accuracy. The resultant velocity vectors of the two groups, taking Galactic rotation into account is presented in Figure~\ref{VelVect}~(right).

The treatments of proper motion velocity vectors presented in Figure~\ref{VelVect} clearly show a marked difference in the proper motions of the two populations, in particular those of the members of L1495 and L1536. In an X-Ray survey \citet{briceno1997x} suggested that a population of stars discovered during the ROSAT mission \citep{trumper1985x}, located to the south of the Taurus clouds, might be an older population and have a different origin from the rest of the cloud, as well as being located at a different distance to the then commonly accepted distance of 140~pc. Our initial findings tend to support these ideas and further suggest that there may be a dynamic link between L1536 and the B213 region within L1495.

Considering Figure~\ref{VelVect}~(right), one star stands out from the rest. Within the `far' population, HD~30067 (Gaia~DR2~147248216395196672) is found to have a markedly different velocity profile to the rest of its group. This star is noted \citet{rebull2010taurus} as being an A4V Class III YSO and is recorded in SIMBAD as being an A2/4 class star located at 163.5$\pm$1.4~pc with a proper motion of $\mu$~=~16.1~mas~yr$^{-1}$. \textit{Gaia} DR2 indicates a G-Band magnitude of 8.9~mag and an extinction of 4.5~mag for this source. We suggest that, although this star meets the distance and proper motion criteria described earlier, HD~30067 is actually a runaway field star and not associated with the Taurus group.

\section{Summary} 
\label{Summary}

This study has shown, through the use of trigonometric parallaxes from \textit{Gaia}~DR2, that there are significant differences in the distances to different structures within the Taurus molecular cloud complex. We have shown that there are two main associations located at 130.6$\pm$0.7 and 160.2$\pm$0.9~pc. The two groups have different proper motions of 24.5$\pm$2.8 and 20.1$\pm$2.4~mas~yr$^{-1}$ respectively, and they appear to be moving in somewhat different directions. They also appear to have slightly different line-of-sight velocities. We call these two populations the `Two Horns' of Taurus.

With this new data we have also been able to confirm that the TMR is not a coherent feature but has an extended depth of approximately 15 parsecs. We also tentatively suggest that the structure of the TMC, in general, resembles that of an inclined sheet facing away from the observer.

\acknowledgments 

We thank the anonymous referees for their constructive comments. Use has been made of data from the European Space Agency (ESA) mission \textit{GAIA} (\href{https://www.cosmos.esa.int/gaia}{https://www.cosmos.esa.int/gaia}), processed by the \textit{GAIA} Data Processing and Analysis Consortium (DPAC) (\href{https://www.cosmos.esa.int/web/gaia/dpac/consortium}{https://www.cosmos.esa.int/web/gaia/dpac/consortium}). Funding for the DPAC has been provided by national institutions, in particular the institutions participating in the \textit{GAIA} Multilateral Agreement.

This work has used the NASA Astrophysics Data System (ADS) Bibliographic Services (\href{http://ads.harvard.edu/}{http://ads.harvard.edu/}) as well as the VizieR catalogue access tool (\href{http://vizier.u-strasbg.fr/viz-bin/VizieR}{http://vizier.u-strasbg.fr/viz-bin/VizieR}) and SIMBAD astronomical database (\href{http://simbad.u-strasbg.fr/simbad/}{http://simbad.u-strasbg.fr/simbad/}), operated at CDS, Strasbourg, France.

Data products have also been used from the Two Micron All Sky Survey, which is a joint project of the University of Massachusetts and the Infrared Processing and Analysis Center/California Institute of Technology, funded by the National Aeronautics and Space Administration and the National Science Foundation. 

This research has made use of {\fontfamily{pcr}\selectfont Astropy}, a community-developed core Python (\href{https://www.python.org/}{https://www.python.org/}) module for Astronomy \citep{robitaille2013astropy, price2018astropy}.  This work has also made extensive use of {\fontfamily{pcr}\selectfont Matplotlib} \citep{hunter2007matplotlib}, {\fontfamily{pcr}\selectfont SciPy} \citep{van2014scikit} and {\fontfamily{pcr}\selectfont NumPy} \citep{van2011numpy}. This work would not have been possible without the countless hours put in by members of the open-source community around the world.

DWT was supported by the UK Science and Technology Facilities Council under grant number ST/R000786/1.
KP acknowledges support from the Ministry of Science and Technology (Taiwan) under Grant No. 106-2119-M-007-021-MY3.  

\bibliographystyle{aasjournal} 
\bibliography{Frodo_Bibliography.bib}

\appendix 
\section{Compendium of Sources}
\label{Compendium}

In this Appendix we list all of the parameters of the 168 sources in our sample, before we applied the final distance and proper motion cuts. The reader may therefore judge the validity of cutting from 168 to 161 sources.

\startlongtable 
\begin{deluxetable*}{LLCCCCCCC}
\tablecaption{Properties of \textit{Gaia} DR2 Taurus sources.\label{chartable}}
\tablecolumns{9}
\tablenum{3}
\tablewidth{0pt}
\tabletypesize{\scriptsize}
\tablehead{
\colhead{\textit{Gaia} DR2 ID} & \colhead{SSTtau ID} & 
\colhead{R.A.} & \colhead{Dec.} & \colhead{Parallax} & \colhead{Distance} & \colhead{PMRA} &\colhead{PMDec} & \colhead{$\mu$} \\
\colhead{} & \colhead{} & \colhead{[deg]} & \colhead{[deg]} & 
\colhead{[mas]} & \colhead{[pc]} & \colhead{[mas yr$^{-1}$]} & 
\colhead{[mas yr$^{-1}$]} & \colhead{[mas yr$^{-1}$]}
} 
\startdata
147869573608324992 & 043051.7+24414 & 67.7156 & 24.6965 & 9.51 & 105.12 & 11.52 & -24.50 & 27.10 \\
151870352825256576 & 043545.2+273713 & 68.9387 & 27.6202 & 8.83 & 113.19 & 14.20 & -25.79 & 29.44 \\
147778490237623808 & 043306.6+240954 & 68.2777 & 24.1652 & 8.83 & 113.20 & 9.56 & -20.23 & 22.37 \\
149685627517927296 & 042359.7+251452 & 65.9988 & 25.2479 & 8.75 & 114.33 & 0.99 & -19.70 & 19.72 \\
147546080967742720 & 043621.5+235116 & 69.0896 & 23.8545 & 8.66 & 115.46 & -1.54 & -19.48 & 19.54 \\
151296579553731456 & 043007.2+260820 & 67.5302 & 26.1390 & 8.62 & 116.06 & 2.67 & -21.927 & 22.09 \\
164513602672978304 & 041840.6+281915 & 64.6692 & 28.3209 & 8.52 & 117.44 & 6.62 & -29.87 & 30.60 \\
164513022853468160 & 041807.9+282603 & 64.5333 & 28.4342 & 8.48 & 117.95 & 8.43 & -24.73 & 26.13 \\
149409920679460096 & 042630.5+244355 & 66.6273 & 24.7321 & 8.41 & 118.91 & 2.74 & -22.02 & 22.19 \\
164502062096975744 & 041901.1+281942 & 64.7546 & 28.3282 & 8.38 & 119.37 & 8.71 & -25.34 & 26.79 \\
164598303725243776 & 041935.4+282721 & 64.8978 & 28.4560 & 8.36 & 119.64 & 11.93 & -25.97 & 28.58 \\
164550882989640192 & 042203.1+282538 & 65.5132 & 28.4274 & 8.31 & 120.27 & 10.27 & -25.86 & 27.83 \\
146874275068113664 & 044000.6+235821 & 70.0028 & 23.9724 & 8.30 & 120.49 & 8.62 & -20.74 & 22.46 \\
151262700852297728 & 042704.6+260616 & 66.7696 & 26.1044 & 8.25 & 121.18 & 6.16 & -19.30 & 20.26 \\
164546038266077824 & 042025.8+281923 & 65.1077 & 28.3232 & 8.19 & 122.13 & 10.66 & -24.22 & 26.46 \\
151102790629500288 & 043057.1+255639 & 67.7384 & 25.9442 & 8.14 & 122.82 & 5.42 & -23.02 & 23.65 \\
148116246465275520 & 044205.4+252256 & 70.5229 & 25.3822 & 8.14 & 122.88 & 5.49 & -18.99 & 19.77 \\
164507353496637952 & 041831.1+281629 & 64.6297 & 28.2747 & 8.12 & 123.08 & 8.82 & -25.64 & 27.12 \\
165563674934601856 & 041357.3+291819 & 63.4891 & 29.3053 & 8.10 & 123.47 & 9.95 & -22.89 & 24.96 \\
146764465639042176 & 043906.3+233417 & 69.7766 & 23.5716 & 8.06 & 124.14 & 8.18 & -21.82 & 23.30 \\
147799209159857280 & 043217.8+242214 & 68.0745 & 24.3707 & 8.05 & 124.17 & 6.26 & -22.20 & 23.06 \\
164519276325850752 & 041817.1+282841 & 64.5713 & 28.4782 & 8.04 & 124.37 & 5.79 & -25.03 & 25.69 \\
146675954953119104 & 043815.6+230227 & 69.5651 & 23.0409 & 7.98 & 125.24 & -1.07 & -17.19 & 17.22 \\
164513400810646912 & 041842.5+281849 & 64.6771 & 28.3138 & 7.97 & 125.50 & 12.13 & -24.15 & 27.02 \\
164495323291866624 & 041851.1+281433 & 64.7132 & 28.2425 & 7.96 & 125.66 & 8.61 & -24.65 & 26.11 \\
164513538249595136 & 041847.0+282007 & 64.6960 & 28.3353 & 7.95 & 125.77 & 8.30 & -25.31 & 26.63 \\
164422961683000320 & 041941.2+274948 & 64.9220 & 27.8299 & 7.91 & 126.37 & 9.94 & -24.96 & 26.87 \\
152118881108855680 & 042445.0+270144 & 66.1878 & 27.0290 & 7.89 & 126.73 & 8.39 & -26.84 & 28.13 \\
146767764173923328 & 043858.5+233635 & 69.7442 & 23.6097 & 7.89 & 126.82 & 8.64 & -21.20 & 22.89 \\
163165738856771200 & 041514.7+280009 & 63.8114 & 28.0025 & 7.88 & 126.88 & 8.55 & -25.52 & 26.92 \\
146881048231272192 & 043933.6+235921 & 69.8902 & 23.9891 & 7.87 & 127.12 & 7.28 & -20.99 & 22.22 \\
152511475478780416 & 042155.6+275506 & 65.4819 & 27.9183 & 7.85 & 127.37 & 10.75 & -27.22 & 29.27 \\
151793082068521856 & 043114.4+271017 & 67.8102 & 27.1715 & 7.84 & 127.49 & 9.17 & -27.58 & 29.06 \\
164445437248152832 & 042026.0+280408 & 65.1087 & 28.0691 & 7.84 & 127.51 & 8.33 & -26.07 & 27.37 \\
146764809236423808 & 043901.6+233602 & 69.7568 & 23.6007 & 7.82 & 127.81 & 8.57 & -21.85 & 23.47 \\
162758236656524416 & 041447.8+264811 & 63.6995 & 26.8030 & 7.82 & 127.93 & 9.02 & -22.45 & 24.20 \\
148037764527442944 & 043619.0+254258 & 69.0796 & 25.7163 & 7.81 & 128.02 & 8.91 & -27.47 & 28.88 \\
163184366130809984 & 041412.9+281212 & 63.5539 & 28.2033 & 7.80 & 128.13 & 8.93 & -29.05 & 30.40 \\
147606657186323712 & 043527.3+241458 & 68.8641 & 24.2496 & 7.80 & 128.22 & 6.05 & -20.77 & 21.64 \\
151327159721125888 & 042945.6+263046 & 67.4403 & 26.5128 & 7.80 & 128.23 & 6.87 & -20.97 & 22.06 \\
147801339463632000 & 043301.9+242100 & 68.2582 & 24.3499 & 7.80 & 128.24 & 6.37 & -20.47 & 21.44 \\
162757545164429696 & 041447.3+264626 & 63.6971 & 26.7739 & 7.78 & 128.47 & 9.48 & -22.69 & 24.59 \\
163246832135164544 & 041314.1+281910 & 63.3090 & 28.3195 & 7.76 & 128.85 & 8.37 & -24.40 & 25.80 \\
164698634160139264 & 041733.7+282046 & 64.3906 & 28.3462 & 7.76 & 128.88 & 9.18 & -25.55 & 27.15 \\
164702070133970944 & 041749.6+282936 & 64.4569 & 28.4933 & 7.75 & 129.03 & 6.90 & -24.62 & 25.57 \\
151373820245230080 & 042920.7+263340 & 67.3363 & 26.5611 & 7.74 & 129.24 & 7.91 & -20.41 & 21.89 \\
149369139966814976 & 042936.0+243555 & 67.4003 & 24.5987 & 7.73 & 129.42 & 8.57 & -20.20 & 21.94 \\
164666022471759232 & 041628.1+280735 & 64.1171 & 28.1265 & 7.70 & 129.84 & 6.85 & -25.90 & 26.80 \\
164409359522965120 & 041830.3+274320 & 64.6263 & 27.7223 & 7.70 & 129.89 & 8.73 & -26.27 & 27.68 \\
145238687096970496 & 043508.5+231139 & 68.7855 & 23.1943 & 7.69 & 129.96 & 7.24 & -20.98 & 22.19 \\
147796013704188928 & 043231.7+242002 & 68.1324 & 24.3341 & 7.69 & 130.03 & 7.12 & -21.50 & 22.65 \\
147796013704189440 & 043230.5+241957 & 68.1274 & 24.3325 & 7.68 & 130.21 & 6.65 & -21.86 & 22.85 \\
146277553787186048 & 043223.2+240301 & 68.0972 & 24.0503 & 7.68 & 130.24 & 7.36 & -22.43 & 23.60 \\
164705368668853120 & 041738.9+283300 & 64.4123 & 28.5500 & 7.68 & 130.27 & 6.87 & -25.25 & 26.17 \\
164536250037820160 & 042158.8+281806 & 65.4952 & 28.3017 & 7.67 & 130.33 & 9.01 & -26.39 & 27.88 \\
151028990206478080 & 043203.2+252807 & 68.0138 & 25.4687 & 7.67 & 130.39 & 8.09 & -22.81 & 24.20 \\
146285112929523456 & 043023.6+235912 & 67.5986 & 23.9868 & 7.67 & 130.40 & 4.73 & -21.97 & 22.48 \\
164518589131083136 & 041831.1+282716 & 64.6297 & 28.4544 & 7.67 & 130.44 & 8.68 & -25.10 & 26.56 \\
147806733942555008 & 043215.4+242859 & 68.0643 & 24.4831 & 7.67 & 130.46 & 5.02 & -21.38 & 21.96 \\
147790202612482560 & 043334.0+242117 & 68.3919 & 24.3547 & 7.66 & 130.50 & 5.88 & -20.55 & 21.38 \\
152226491513195648 & 042457.0+271156 & 66.2379 & 27.1989 & 7.66 & 130.57 & 8.36 & -26.75 & 28.02 \\
147831571737487488 & 043310.0+243343 & 68.2918 & 24.5619 & 7.66 & 130.60 & 7.25 & -21.23 & 22.43 \\
163182888662060928 & 041411.8+281153 & 63.5496 & 28.1981 & 7.63 & 131.09 & 9.08 & -23.89 & 25.56 \\
146366442430208640 & 042959.5+243307 & 67.4980 & 24.5520 & 7.63 & 131.12 & 8.18 & -21.44 & 22.95 \\
148017561002336384 & 043342.9+252647 & 68.4289 & 25.4462 & 7.63 & 131.14 & 5.13 & -22.77 & 23.35 \\
164738521519622656 & 041414.5+282758 & 63.5609 & 28.4660 & 7.62 & 131.24 & 9.50 & -23.57 & 25.41 \\
151297958238753664 & 042951.5+260644 & 67.4648 & 26.1124 & 7.62 & 131.28 & 5.95 & -21.01 & 21.84 \\
150393571269837184 & 041810.7+251957 & 64.5450 & 25.3325 & 7.61 & 131.36 & 8.83 & -23.34 & 24.95 \\
164684340508950144 & 041539.1+281858 & 63.9132 & 28.3162 & 7.61 & 131.49 & 9.25 & -24.38 & 26.08 \\
164504467278644096 & 041926.2+282614 & 64.8595 & 28.4372 & 7.59 & 131.70 & 8.36 & -25.47 & 26.81 \\
163184366130809472 & 041413.5+281249 & 63.5566 & 28.2136 & 7.58 & 131.94 & 8.59 & -24.42 & 25.88 \\
163184091252903936 & 041417.0+281057 & 63.5709 & 28.1826 & 7.55 & 132.44 & 8.28 & -24.16 & 25.54 \\
152917298349085824 & 042515.5+282927 & 66.3147 & 28.4909 & 7.52 & 133.06 & 10.79 & -25.05 & 27.27 \\
152362491654557696 & 042039.1+271731 & 65.1633 & 27.2920 & 7.51 & 133.07 & 10.41 & -26.09 & 28.09 \\
147847072275324416 & 043150.5+242418 & 67.9607 & 24.4048 & 7.50 & 133.33 & 4.46 & -22.96 & 23.39 \\
151262876946558976 & 042657.3+260628 & 66.7388 & 26.1078 & 7.50 & 133.36 & 5.19 & -20.71 & 21.35 \\
151125919028356352 & 043336.7+260949 & 68.4033 & 26.1636 & 7.50 & 133.40 & 8.16 & -17.37 & 19.19 \\
163179521407696384 & 041505.1+280846 & 63.7715 & 28.1460 & 7.49 & 133.57 & 8.46 & -24.48 & 25.90 \\
152516079683687680 & 042306.0+280119 & 65.7754 & 28.0220 & 7.47 & 133.88 & 7.93 & -26.68 & 27.84 \\
148450085683504896 & 043835.2+261038 & 69.6471 & 26.1773 & 7.46 & 133.96 & 4.89 & -21.78 & 22.32 \\
147869784062378624 & 043051.3+244222 & 67.7141 & 24.7061 & 7.45 & 134.22 & 7.05 & -20.94 & 22.10 \\
151265002954775936 & 042727.9+261205 & 66.8667 & 26.2013 & 7.42 & 134.69 & 1.81 & -23.45 & 23.52 \\
164514053645658752 & 041901.9+282233 & 64.7583 & 28.3758 & 7.40 & 135.07 & 9.19 & -25.76 & 27.35 \\
151374198202645376 & 042942.4+263249 & 67.4270 & 26.5469 & 7.40 & 135.12 & 6.90 & -21.21 & 22.30 \\
162967384383246336 & 041557.9+274617 & 63.9917 & 27.7714 & 7.37 & 135.67 & 7.82 & -24.65 & 25.86 \\
163181342473839744 & 041417.6+28060 & 63.5734 & 28.1026 & 7.37 & 135.69 & 8.34 & -23.32 & 24.76 \\
148172179824515968 & 044148.2+253430 & 70.4511 & 25.5751 & 7.34 & 136.16 & 4.51 & -19.61 & 20.12 \\
148112913570653568 & 044221.0+252034 & 70.5876 & 25.3428 & 7.34 & 136.16 & 4.89 & -19.42 & 20.03 \\
163233981593016064 & 041327.2+281624 & 63.3635 & 28.2734 & 7.34 & 136.30 & 7.44 & -23.84 & 24.97 \\
148401565437820928 & 044008.0+260525 & 70.0334 & 26.0903 & 7.34 & 136.32 & 5.73 & -20.13 & 20.93 \\
148400229703257856 & 043944.8+260152 & 69.9370 & 26.0312 & 7.31 & 136.85 & 7.46 & -22.11 & 23.33 \\
153001307909276928 & 042916.2+285627 & 67.3176 & 28.9409 & 7.30 & 137.05 & -6.78 & 5.99 & 9.05 \\
147818450613367424 & 043455.4+242853 & 68.7309 & 24.4813 & 7.29 & 137.20 & 3.48 & -20.99 & 21.27 \\
164470794735041152 & 041618.8+275215 & 64.0786 & 27.8708 & 7.28 & 137.29 & 7.47 & -24.93 & 26.02 \\
164506116546058112 & 041749.5+281331 & 64.4565 & 28.2254 & 7.28 & 137.37 & 8.81 & -25.22 & 26.71 \\
164474986623118592 & 041612.1+275638 & 64.0505 & 27.9439 & 7.27 & 137.49 & 8.89 & -25.72 & 27.21 \\
152029992465874560 & 042420.9+263051 & 66.0871 & 26.5141 & 7.27 & 137.59 & 9.25 & -27.37 & 28.89 \\
148449845165337600 & 043821.3+260913 & 69.5889 & 26.1537 & 7.23 & 138.31 & 5.47 & -22.97 & 23.62 \\
152518828462749440 & 042307.7+280557 & 65.7824 & 28.0992 & 7.19 & 139.05 & 9.65 & -27.01 & 28.69 \\
148449913884294528 & 043828.5+261049 & 69.6191 & 26.1803 & 7.17 & 139.38 & 6.13 & -21.34 & 22.20 \\
151262941369626752 & 042654.4+260651 & 66.7267 & 26.1141 & 7.15 & 139.91 & 3.98 & -20.40 & 20.79 \\
145157937416226176 & 043559.4+223829 & 68.9980 & 22.6413 & 7.13 & 140.33 & 11.85 & -16.94 & 20.67 \\
148384179410294272 & 044108.2+255607 & 70.2845 & 25.9353 & 7.10 & 140.83 & 7.20 & -21.95 & 23.10 \\
147441558642852736 & 044427.1+251216 & 71.1131 & 25.2045 & 7.09 & 141.01 & 6.45 & -20.15 & 21.16 \\
147679014500233728 & 044018.8+243234 & 70.0786 & 24.5427 & 7.07 & 141.50 & -1.41 & -43.35 & 43.38 \\
163181308112262400 & 041426.2+280603 & 63.6095 & 28.1008 & 7.04 & 142.13 & 5.54 & -27.64 & 28.19 \\
151787064819255936 & 043126.6+270318 & 67.8613 & 27.0551 & 6.99 & 143.08 & 13.99 & -19.89 & 24.32 \\
148354733113981696 & 043903.9+254426 & 69.7665 & 25.7406 & 6.95 & 143.99 & 7.04 & -20.61 & 21.77 \\
148106316500918272 & 044303.0+252018 & 70.7628 & 25.3384 & 6.92 & 144.57 & 4.73 & -20.21 & 20.75 \\
163229544890946944 & 041353.2+281123 & 63.4721 & 28.1897 & 6.91 & 144.80 & 11.32 & -22.68 & 25.35 \\
148450875956969344 & 043814.8+261139 & 69.5620 & 26.1943 & 6.88 & 145.41 & 4.01 & -23.17 & 23.51 \\
148420639387738112 & 043800.8+255857 & 69.5036 & 25.9825 & 6.87 & 145.50 & 4.90 & -22.63 & 23.16 \\
152643240779301632 & 042900.6+275503 & 67.2529 & 27.9175 & 6.87 & 145.65 & 8.65 & -25.26 & 26.70 \\
145196527698016512 & 043319.0+224634 & 68.3295 & 22.7761 & 6.77 & 147.67 & 10.94 & -20.15 & 22.93 \\
145947077527182848 & 043309.4+224648 & 68.2895 & 22.7801 & 6.71 & 148.98 & 10.67 & -16.76 & 19.87 \\
148374391180009600 & 043955.7+254502 & 69.9823 & 25.7505 & 6.70 & 149.27 & 6.19 & -20.34 & 21.26 \\
150501362066641664 & 042216.4+254911 & 65.5686 & 25.8199 & 6.61 & 151.26 & 14.13 & -19.65 & 24.20 \\
148141775750936960 & 044039.7+251906 & 70.1658 & 25.3183 & 6.57 & 152.20 & 6.35 & -20.57 & 21.53 \\
151130591952773632 & 043307.8+261606 & 68.2826 & 26.2684 & 6.57 & 152.25 & 7.17 & -17.31 & 18.74 \\
151037064744973696 & 043158.4+254329 & 67.9936 & 25.7249 & 6.55 & 152.60 & 7.55 & -21.14 & 22.45 \\
145133786815830784 & 043541.8+223411 & 68.9244 & 22.5698 & 6.53 & 153.25 & 10.96 & -17.96 & 21.04 \\
147373010964871040 & 044642.6+245903 & 71.6776 & 24.9842 & 6.48 & 154.25 & 4.60 & -19.52 & 20.06 \\
152109054223716480 & 042423.2+265008 & 66.0968 & 26.8356 & 6.46 & 154.73 & 11.49 & -18.35 & 21.65 \\
145950895754320384 & 043224.1+225108 & 68.1007 & 22.8522 & 6.44 & 155.36 & 9.99 & -17.43 & 20.09 \\
152108882425024128 & 042426.4+264950 & 66.1103 & 26.8305 & 6.43 & 155.51 & 11.58 & -17.86 & 21.29 \\
151283870746458496 & 042629.3+262413 & 66.6225 & 26.4037 & 6.42 & 155.88 & 10.90 & -17.85 & 20.92 \\
145209442664192896 & 043552.8+225058 & 68.9703 & 22.8495 & 6.39 & 156.53 & 10.95 & -16.53 & 19.83 \\
148196510814073728 & 044110.7+255511 & 70.2950 & 25.9198 & 6.38 & 156.81 & 4.52 & -20.11 & 20.61 \\
163183644576299264 & 041449.2+281230 & 63.7054 & 28.2084 & 6.37 & 156.92 & 10.65 & -21.17 & 23.70 \\
145213192171160064 & 043553.4+225408 & 68.9730 & 22.9024 & 6.36 & 157.15 & 11.37 & -18.48 & 21.70 \\
152288824375681536 & 042216.7+265457 & 65.5699 & 26.9158 & 6.35 & 157.57 & 11.31 & -17.60 & 20.92 \\
152098299625634816 & 042247.8+264553 & 65.6996 & 26.7646 & 6.34 & 157.67 & 10.74 & -17.14 & 20.23 \\
145209618758377856 & 043547.3+225021 & 68.9473 & 22.8393 & 6.30 & 158.63 & 13.52 & -15.90 & 20.87 \\
145203159127518336 & 043352.0+225030 & 68.4668 & 22.8416 & 6.30 & 158.71 & 8.90 & -17.07 & 19.25 \\
144936836795636864 & 043917.7+222103 & 69.8242 & 22.3509 & 6.29 & 158.87 & 10.47 & -17.38 & 20.29 \\
145157941711889536 & 043558.9+223835 & 68.9956 & 22.6431 & 6.29 & 158.94 & 10.81 & -15.77 & 19.12 \\
145196252820109440 & 043326.2+224529 & 68.3593 & 22.7581 & 6.28 & 159.17 & 8.62 & -16.75 & 18.84 \\
145213295250374016 & 043552.0+225503 & 68.9671 & 22.9177 & 6.28 & 159.22 & 8.78 & -8.09 & 11.94 \\
148010281032823552 & 043339.0+252038 & 68.4129 & 25.3438 & 6.28 & 159.34 & 9.33 & -18.29 & 20.53 \\
152000786688289664 & 042444.5+261014 & 66.1858 & 26.1705 & 6.27 & 159.57 & 11.87 & -17.32 & 21.00 \\
148289831863907840 & 044138.8+255626 & 70.4118 & 25.9407 & 6.26 & 159.78 & 2.50 & -15.48 & 15.68 \\
164800235906366976 & 041542.7+290959 & 63.9283 & 29.1665 & 6.25 & 159.95 & 12.34 & -17.75 & 21.62 \\
164783811951433856 & 041639.1+285849 & 64.1631 & 28.9803 & 6.25 & 160.04 & 4.97 & -15.00 & 15.80 \\
148116276529733120 & 044207.7+252311 & 70.5325 & 25.3866 & 6.25 & 160.12 & -0.74 & -14.15 & 14.17 \\
146319193494413696 & 042745.3+235724 & 66.9392 & 23.9567 & 6.24 & 160.18 & 11.27 & -16.57 & 20.04 \\
152293149405058816 & 042146.3+265929 & 65.4430 & 26.9914 & 6.23 & 160.47 & 11.92 & -18.33 & 21.86 \\
152104381299305600 & 042449.0+264310 & 66.2044 & 26.7195 & 6.21 & 160.97 & 10.61 & -17.05 & 20.08 \\
145132927822383616 & 043520.2+223214 & 68.8343 & 22.5373 & 6.20 & 161.40 & 9.83 & -16.63 & 19.32 \\
145203811962545152 & 043410.9+225144 & 68.5459 & 22.8623 & 6.18 & 161.77 & 8.90 & -17.27 & 19.43 \\
145213875069914496 & 043520.8+225424 & 68.8371 & 22.9066 & 6.18 & 161.83 & 9.72 & -19.25 & 21.56 \\
145212711134828672 & 043542.0+225222 & 68.9252 & 22.8729 & 6.17 & 162.07 & 10.02 & -16.77 & 19.54 \\
145210099794710272 & 043551.0+225240 & 68.9629 & 22.8777 & 6.16 & 162.32 & 6.32 & -28.15 & 28.85 \\
145225596036660224 & 043456.9+225835 & 68.7373 & 22.9765 & 6.15 & 162.68 & 11.70 & -16.41 & 20.16 \\
147562470562750720 & 043649.1+241258 & 69.2049 & 24.2163 & 6.14 & 162.85 & 43.42 & -13.26 & 45.40 \\
152099055539792000 & 042224.0+264625 & 65.6002 & 26.7738 & 6.13 & 163.27 & 11.41 & -17.86 & 21.19 \\
147248216395196672 & 044518.2+242436 & 71.3258 & 24.4101 & 6.11 & 163.54 & -15.85 & -2.85 & 16.10 \\
145951789107603200 & 043249.1+225302 & 68.2047 & 22.8841 & 6.08 & 164.53 & 7.43 & -15.79 & 17.45 \\
145213192171159552 & 043554.1+225413 & 68.9757 & 22.9037 & 6.03 & 165.95 & 11.87 & -9.99 & 15.51 \\
152305248330621184 & 042134.5+270138 & 65.3942 & 27.0273 & 5.99 & 167.00 & 11.77 & -16.63 & 20.38 \\
164800815725933312 & 041524.0+291043 & 63.8505 & 29.1787 & 5.97 & 167.54 & 10.74 & -19.44 & 22.20 \\
152349022637314176 & 042025.5+270035 & 65.1065 & 27.0098 & 5.87 & 170.35 & 11.35 & -17.73 & 21.05 \\
145220064117853696 & 043610.3+225956 & 69.0433 & 22.9988 & 5.84 & 171.24 & 10.81 & -15.96 & 19.27 \\
146050057959093632 & 043119.0+233504 & 67.8295 & 23.5846 & 5.83 & 171.56 & 8.08 & -16.60 & 18.47 \\
145203704587705088 & 043415.2+225030 & 68.5637 & 22.8418 & 5.82 & 171.97 & 10.08 & -17.50 & 20.19 \\
151129011404806912 & 043344.6+261500 & 68.4361 & 26.2500 & 5.77 & 173.26 & 6.51 & -17.31 & 18.50 \\
145213187879627776 & 043552.7+225423 & 68.9700 & 22.9064 & 5.64 & 177.15 & 8.82 & -13.95 & 16.51 \\
145217379763796992 & 043638.9+225811 & 69.1622 & 22.9699 & 5.44 & 183.91 & 9.55 & -15.97 & 18.61 \\
152361426502650496 & 042115.2+272101 & 65.3136 & 27.3502 & 5.41 & 184.86 & 2.96 & -4.53 & 5.41 \\
\enddata
\end{deluxetable*}

\end{document}